\documentclass[journal]{IEEEtran}
\ifCLASSINFOpdf
\else
\fi
\usepackage{graphicx}
\usepackage{comment}
\usepackage{amsmath,amssymb} 
\usepackage{color}
\usepackage{url}

\usepackage{amsfonts}  

\usepackage{epsfig}
\usepackage{graphicx}
\usepackage{subcaption}
\usepackage{amsmath}
\usepackage{amssymb}
\usepackage{xcolor}
\usepackage{wrapfig}
\usepackage{xspace}
\usepackage{array, multirow}
\usepackage{upgreek}
\usepackage{makecell}

\usepackage[nolist,nohyperlinks]{acronym}

\usepackage{easylist}

\newcommand{\ie}{\textit{i}.\textit{e}.}

\DeclareMathOperator*{\argmax}{argmax}

\acrodef{PSNR}[PSNR]{{peak signal to noise ratio}}
\acrodef{SSIM}[SSIM]{{structural similarity index measure}}
\acrodef{HDR}[HDR]{{high-dynamic-range}}
\acrodef{SDR}[SDR]{{standard-dynamic-range}}
\acrodef{IQA}[IQA]{{image quality assessment}}
\acrodef{PU}[PU]{{perceptually uniform}}
\acrodef{SROCC}[SROCC]{{spearman rank order correlation coefficient}}
\acrodef{HDR-VDP}[HDR-VDP]{{HDR visual difference predictor}}
\acrodef{HDR-VQM}[HDR-VQM]{{HDR video quality measure}}
\acrodef{CRT}[CRT]{{cathode ray tube}}
\acrodef{PLCC}[PLCC]{{pearson linear correlation coefficient}}
\acrodef{CNN}[CNN]{{convolutional neural network}}
\acrodef{RMSE}[RMSE]{{root-mean-squared error}}
\acrodef{ML}[ML]{{machine learning}}
\acrodef{MOS}[MOS]{{Mean Opinion Scores}}

\newcommand{\cdms}{\,cd/m$^2$\xspace}

\begin{document}
%
\title{Consolidated Dataset and Metrics \\ for High-Dynamic-Range Image Quality}
%
%

\author{Aliaksei~Mikhailiuk, Mar\'ia P\'erez-Ortiz, Dingcheng~Yue, Wilson Suen, and Rafa{\l} K. Mantiuk
\thanks{A. Mikhailiuk, D. Yue, W. Suen and R. Mantiuk are with the Department of Computer Science and Technology at the University of Cambridge (UK) (email: \{am2442, dy276, wss28, rkm38\}@cam.ac.uk).} 
\thanks{M. P\'erez-Ortiz is with the Department of Computer Science at the University College London (UK) (email: maria.perez@ucl.ac.uk)}
}

%
%

\markboth{Journal of \LaTeX\ Class Files,~Vol.~14, No.~8, August~2015}%
{Shell \MakeLowercase{\textit{et al.}}: Bare Demo of IEEEtran.cls for IEEE Journals}
%



\maketitle

\begin{abstract}
Increasing popularity of \ac{HDR} image and video content brings the need for metrics that could predict the severity of image impairments as seen on displays of different brightness levels and dynamic range. Such metrics should be trained and validated on a sufficiently large subjective image quality dataset to ensure robust performance. As the existing \ac{HDR} quality datasets are limited in size, we created a Unified Photometric Image Quality dataset (UPIQ) with over 4,000 images by realigning and merging existing \ac{HDR} and \ac{SDR} datasets. The realigned quality scores share the same unified quality scale across all datasets. Such realignment was achieved by collecting additional cross-dataset quality comparisons and re-scaling data with a psychometric scaling method. Images in the proposed dataset are represented in absolute photometric and colorimetric units, corresponding to light emitted from a display. We use the new dataset to retrain existing HDR metrics and show that the dataset is sufficiently large for training deep architectures. We show the utility of the dataset on brightness aware image compression.
\end{abstract}

\begin{IEEEkeywords}
High Dynamic Range, Image Quality Dataset, Image Quality Metric
\end{IEEEkeywords}

\IEEEpeerreviewmaketitle

\section{Introduction}
%
%
%
%
\IEEEPARstart{I}{mage} quality assessment metrics, such as \ac{PSNR} and \ac{SSIM} are widely used in image compression, reconstruction, and enhancement \cite{DBLP:journals/corr/abs-1709-05424,7797130,4271520,7949028,bovik_book}. However, most \ac{IQA} metrics do not account for display characteristics such as the dynamic range and brightness of the display, influencing the perceived image quality. For example, compression artifacts are more visible on a bright \ac{HDR} display, than on a dimmed mobile phone \cite{photometricnn2019}. The plethora of display types motivates the need for a new, \textit{photometric} \ac{IQA} metric that accounts for absolute image luminance and can operate on both \ac{HDR} and \ac{SDR} images. {Throughout this work we use the term \emph{metric} to refer to image quality metric rather than to a distance function in a metric space.}

The primary limitation to developing an HDR image quality metric has been the lack of a unified large-scale subjective image quality dataset. Although attempts have been made to adapt and verify performance of SDR metrics on HDR content \cite{Aydn2008,peooqmfhic,etpoefrqmohdrvc,Hanhart2015,Zerman2017_allignment} 
those have not been thoroughly tested due to the lack of a unified dataset. The absence of a large unified dataset also prevented the development of metrics based on machine learning for HDR images, which require large amounts of versatile and heterogeneous data to train. 
While recent machine-learning-based \ac{SDR} image quality metrics relied on large crowd-sourcing studies \cite{Prashnani_2018_CVPR,zhang2018perceptual,Hosu_2020}, these are not straight forward to conduct for \ac{HDR} content as it requires an \ac{HDR} display and controlled viewing conditions.

The available subjective image quality datasets \cite{Ponomarenko2015,Narwaria2013,Zerman2017_allignment,Sheikh2006b,korshunov2015,6489321,Prashnani_2018_CVPR,zhang2018perceptual,7327186}, are insufficient in isolation, as they are limited in terms of the number of images, diversity of distortion types and image sizes. These datasets cannot be easily combined, due to the use of different experimental protocols and the relative nature of the quality scales, which precludes comparing quality scores across datasets. Moreover, incomparable quality scales across datasets prevent the use of absolute scores as a mean of benchmarking \ac{IQA} metrics, forcing to rely on correlation coefficients computed individually on each dataset. 
This work addresses these issues. Instead of following the common practice of collecting a dataset from scratch, we argue for consolidation of existing datasets and focus on combining \ac{SDR} and \ac{HDR} image quality datasets to create the largest photometric subjective \ac{IQA} dataset to date with a unified quality scale. We use the dataset to re-train existing full-reference metrics, including deep architectures.

Our contributions can be summarized as follows: (i) we perform a series of subjective image quality assessment experiments and construct the largest subjective HDR IQA dataset to date (UPIQ) using psychometric scaling \cite{2019TIP}. The dataset contains \textit{3779} SDR and \textit{380} HDR images from four existing IQA datasets; (ii) we show the necessity and advantages of the psychometric scaling by comparing it to other strategies for merging datasets; (iii) we use the new dataset to retrain and benchmark existing HDR metrics. We show that the proposed dataset is sufficiently large for deep architectures by training a \ac{CNN}-based full-reference \textit{photometric} image quality metric. 
The advantage of training on the unified dataset is shown in comparison with training on a single dataset and performing multi-task learning on disjoint datasets; (iv) the utility of training HDR metrics on the new dataset is shown in an application to image compression. The new dataset\footnote{UPIQ dataset: \url{https://doi.org/10.17863/CAM.68372}}, code and metrics are available online\footnote{Project page: \url{https://www.cl.cam.ac.uk/research/rainbow/projects/upiq/}}.

\section{Related Work}\label{sec:related_work}
In this section we set the context for the problem and provide a brief review on existing IQA datasets as well as the HDR and SDR objective quality metrics.

\subsection{Existing IQA Datasets} \label{sec:datasets}

To train and validate image quality metrics, one requires a dataset where image quality scores are obtained from human observer judgements. 
Two most common experimental protocols for collecting such a dataset are ranking and rating experiments. In rating experiments, observers are asked to assign a numeric quality score to each image. All judgments are then averaged to produce \ac{MOS}. In ranking, observers are asked to compare two or more images and order them according to their quality. Ranking results can be then mapped to a one-dimensional quality scale using psychometric scaling \cite{NIPS2006_3079,BradleyTerry,Thurstone1927-THUALO-2,perez2017}. The most commonly used ranking protocol is pairwise comparison, where a pair of images is compared at a time. Advantages of pairwise comparisons over rating are the lower cognitive load on observers and a more accurate scale \cite{Zerman2018}. Unlike rating, scores produced by psychometric scaling are interpretable -- distance between any two scores can be mapped to the probability of one condition being selected over another \cite{perez2017}.

Although many subjective IQA datasets exist, they are far from ideal. For example, the largest currently available \ac{SDR} dataset, BAPPS \cite{zhang2018perceptual}, offers only a single distortion type per content (where we define content as the scene depicted in the image) --- therefore machine learning based metrics may struggle to learn how to scale the magnitude of a distortion. Moreover, image quality scores were not measured extensively, with only two judgments per $64\times64$ pixel patches rather than full-sized images. Another recently collected large-scale \ac{SDR} dataset \cite{Prashnani_2018_CVPR}, contains pairwise preference probability, \ie~the likelihood of an image in a pair of being more similar to the reference. Even though authors collected a large number of comparisons per image ($>10$), only within-content comparisons were performed, thus making the cross-content quality scores less reliable. Similarly authors in \cite{kadid10k} collected a large scale \ac{SDR} dataset with \ac{MOS}. These datasets are not publicly available. Existing \ac{HDR} IQA datasets \cite{Zerman2017_allignment,korshunov2015,Narwaria2013} are significantly smaller than \ac{SDR} datasets, more homogenous in the versatility of their contents and distortions. These datasets are thus insufficient for the applications outlined in this paper. To remedy these limitations, in this work we combine \ac{HDR} datasets with much larger \ac{SDR} datasets.

Since collecting large amount of \ac{IQA} measurements is time consuming and costly, it is preferable to reuse existing datasets. The idea of combining subjective \ac{IQA} datasets has been considered before. Authors in \cite{Zerman2017_allignment} align subjective scores of HDR datasets using objective quality metrics. The method assumes that the quality predictions from multiple objective metrics can be used to find the transformation of quality scores from one dataset to another. However, this approach is problematic when combining \ac{SDR} and \ac{HDR} datasets as very few metrics can reliably predict absolute quality of both \ac{SDR} and \ac{HDR} images. In contrast to that work, we conduct a set of subjective experiments to measure the relative cross-dataset quality and then use psychometric scaling \cite{2019TIP} procedure to bring all datasets to a unified quality scale.

Datasets can be collected more efficiently using \emph{active sampling} methods, which choose the measurements that maximize the information gain \cite{ye2014active,Mikhailiuk2020}. Another approach is to use quality metrics to find the conditions for which the metrics disagree the most and help to differentiate the performance of those metrics (MAD competition) \cite{zhou2008,Ma2020}. Those methods, however, are not intended for merging the existing large datasets.

\subsection{Quality Assessment Criteria}

There are at least three common criteria related to image quality: aesthetics, visibility and impairment. Aesthetic judgements deal with the quality of an image as judged by commonly established photographic rules --- appropriate use of lighting, contrast, and image composition. Here, the quality may be perceived in terms of creative composition and execution of an image, rather than artifacts \cite{7974874}. As an example, tone-mapping metrics, which assess the reproduction of \ac{HDR} images on regular monitors, estimate the aesthetics of tones, brightness, details and colors \cite{Hadizadeh2018,Krasula2020}. {Visibility metrics} predict whether a difference between a pair of images is going to be visible, but they do not assess the magnitude of a distortion \cite{Mantiuk:2011:HCV:2010324.1964935,Wolski2018}. They also produce visual difference maps rather than a single quality score. The focus of this work are impairment metrics, which assess the quality of images distorted by noise, blur, compression, and other artifacts. Here, we only consider full-reference metrics, for which the original undistorted image is available.

\subsection{Existing Approaches to IQA}

Metrics vary in the number of trainable parameters and therefore the amount of data required to train them. The simplest metrics, such as PSNR and SSIM \cite{1284395}, are designed to capture image statistics that is deemed to be important for detecting distortions. These metrics do not have any trainable parameters. Other metrics involve modeling relevant characteristic of visual system, such as contrast masking or contrast sensitivity \cite{NARWARIA201546,Mantiuk:2011:HCV:2010324.1964935}. Because those metrics rely on the existing psychophysical models, they have only a few parameters to train. Another common approach is to extract a number of hand-crafted features and then use a machine-learning model to predict quality based on those \cite{Li2016:VMAF,6272356,7084843,Mittal2013MakingA}. The last group of methods relies on deep-learning methods to both learn the features relevant for quality and the function that would map those features into image quality \cite{dlohvsiiqaf,GAO2017104,zhang2018perceptual,Prashnani_2018_CVPR,8103112,DBLP:journals/corr/LiuWB17}. Deep learning methods have achieved the state-of-the-art performance on several benchmarks, however, they are susceptible to the quality and quantity of the training data. If data are scarce, which is common in IQA, given the difficulty in collecting the  human judgements, the model will fail to generalize.

\subsection{Training Robust Deep Learning models}

Transfer learning is often used to alleviate the problem of insufficient and noisy data when training deep-learning models. For example, authors in \cite{GAO2017104,iqabccfbi,7868646,Bianco2018} pre-trained a \ac{CNN} model on image classification tasks, arguing that learned features would capture image statistics important for IQA. 
Others \cite{zhang2018perceptual,deepfl-iqa} pre-trained the network on the quality predictions of the hand-crafted quality metrics, and then fine-tuned on the subjective image quality scores. 
Authors in \cite{ma2018} exploited yet another approach --- they first pre-trained the network to classify distortions. The risk of this approach, however, is that it can overfit the model to the given set of distortions. Machine learning metrics can also be trained to rank pairs of images rather than rating them \cite{DBLP:journals/corr/LiuWB17,Prashnani_2018_CVPR,zhang2018perceptual}. The advantage of such approach is that training can be performed directly on the pairwise comparison data. However, the shortcoming of this approach is that it discards meaningful information by converting the quality scores to a binary classification problem. Although all these approaches can improve the ability of the ML-metrics to generalize, they do not alleviate the need for a larger and diverse IQA dataset, an issue widely acknowledged in most works \cite{8540075}.

\subsection{Influence of a Display on the Perceived Image Quality}
\label{sec:PU-encoding}
SDR metrics typically operate on 8-bit gamma-encoded pixel values and ignore display characteristics, such as its brightness and resolution and viewing conditions such as viewing distance. Such an approach was justifiable in the era of \ac{CRT} monitors with very similar characteristic. However, current display devices can vary widely. For example, peak display luminance can vary from 5\cdms for a dimmed mobile phone to 6\,000\cdms for a bright HDR display. {One way to account for display brightness is to represent both \ac{SDR} and \ac{HDR} images in absolute colorimetric units, so that, for example RGB=[100\ 100\ 100], corresponds to white color (D65) of 100\cdms. This changes the paradigm of assessing image quality from device-independent measurements (e.g. PSNR on gamma-encoded pixels), to device-specific measurements, which require knowledge of the target display. While this introduces additional difficulty of selecting display parameters, it is a necessary step for quality assessment on modern displays, and especially those with HDR capabilities. The standardization of reference display parameters may simplify this step in the future.}

\subsubsection{Display model} Most modern displays can be modelled using a gain-gamma-offset model \cite{Hainich2016PerceptualDC}. Such a model transforms gamma-encoded color values into absolute linear color values as follows: 
\begin{equation}\label{eq:disp_model_SDR}
       C_{\mathrm{lin}} = (L_{\mathrm{peak}}-L_{\mathrm{black}})\,C_{\mathrm{sRGB}}^\gamma + L_{\mathrm{black}},
\end{equation}
where $\gamma=2.2$, $C_{\mathrm{lin}}$ is a linear color value, $C_{\mathrm{sRGB}}$ is the gamma-encoded color value for one of the channels (R, G or B) and $L_{\mathrm{peak}}$ and $L_{\mathrm{black}}$ are peak and black luminance levels of the display, respectively.

\subsubsection{Photometric image quality metrics} As the dynamic range of a display affects the visibility of distortions of a viewed image, a reliable quality metric should be able to account for it. We will refer to the metrics that operate on physical photometric/luminance values as \emph{photometric quality metrics}. \ac{HDR} quality metrics, such as \ac{HDR-VDP}~\cite{Mantiuk:2011:HCV:2010324.1964935} or \ac{HDR-VQM}~\cite{NARWARIA201546}) are photometric and account for a large range of luminance produced by \ac{HDR} displays. 

\subsubsection{Extending SRD metrics to HDR images}\ac{SDR} metrics can also be adapted to operate on photometric quantities \cite{Mantiuk2016a}. For that, the absolute luminance values are converted into Perceptually Uniform (PU) or logarithmic values, with the former achieving better results \cite{Aydn2008,korshunov2015}. This transformation is necessary as the response of the human eye to luminance is not linear.  Perceptually uniform values are then passed to an SDR image quality metric.

\begin{figure}[t]
\centering
\includegraphics[width=\linewidth]{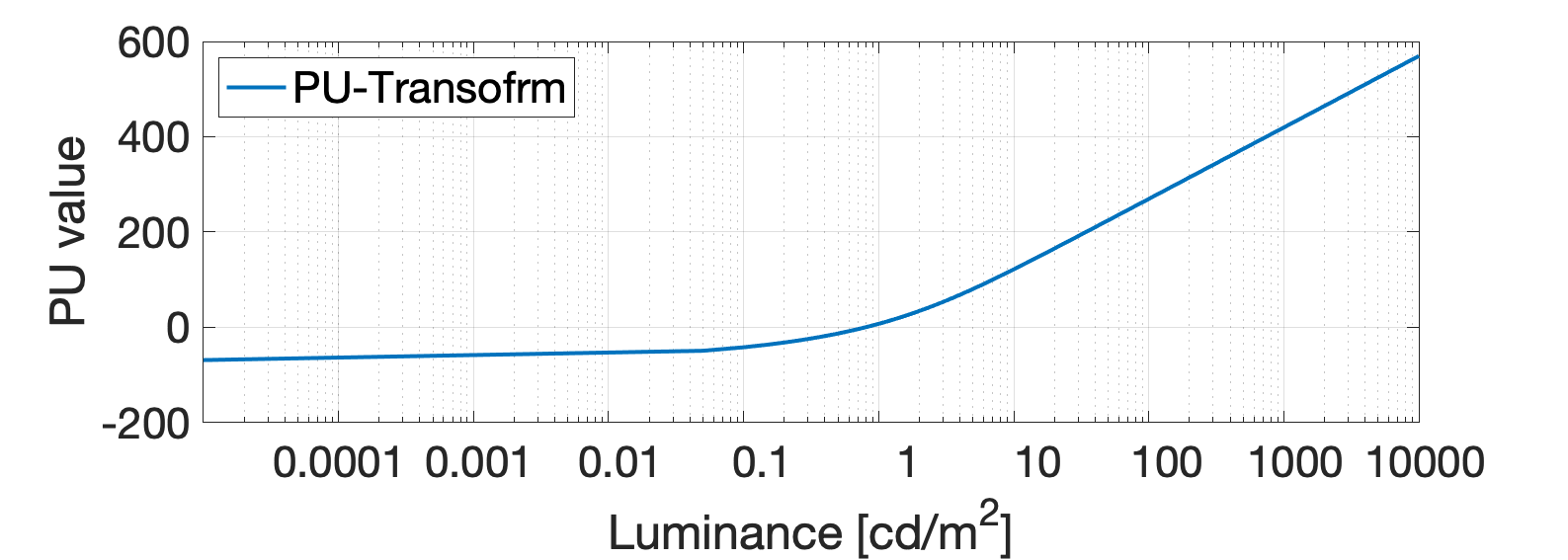}
    \caption{PU-transform mapping physical luminance in \cdms into approximately perceptually uniform units.}\label{fig:pu_transform}
\end{figure}

PU-transform \cite{Aydn2008} was derived from the contrast sensitivity function (CSF) that predicts detection thresholds of the human visual system across a broad range of luminance adaptation conditions.  The transform was designed to ensure that the smallest perceivable change in luminance (just-noticeable-difference or detection threshold) is mapped to a constant change in the PU values. This was achieved by a numerical solution of: 
 \begin{equation}
    PU(L) = \int_{L_{min}}^{L} \frac{1}{T(l)}dl\,,
\end{equation}
where $L_{min}$ is the minimum luminance to be encoded and $T(l)$ is the detection threshold at the absolute luminance $l$. The shape of PU-transform is shown in Figure \ref{fig:pu_transform}. The transformation is typically stored as a look-up table. 
The transformation is further constrained to map luminance values typically reproduced on SDR displays (0.8-80 \cdms) to $0-255$ range (8-bit encoding). This ensures that metric predictions for SDR images in the PU domain are comparable to those computed on the original SDR pixel values.

\begin{figure}[t]
\centering
\includegraphics[width=\linewidth]{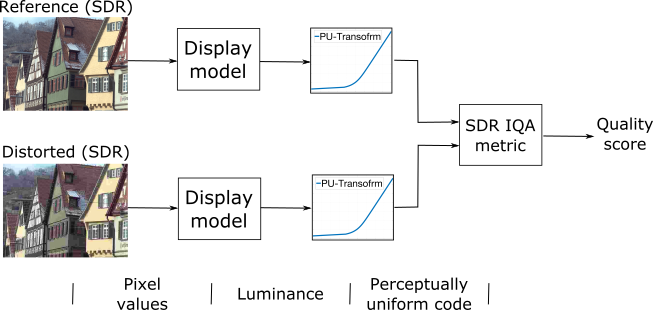}
    \caption{Extending SDR metrics to photometric values. An SDR image is first converted from gamma-corrected pixel values to linear luminance values, via a display model. For HDR images the ``Display model'' is omitted, as those already store luminance values. Luminance values are then passed via the PU-transform to obtain perceptually uniform values. The images in PU domain are then passed to an SDR image quality metric.} \label{fig:extending_sdr_to_hdr}
\end{figure}

Similar procedure can be applied to evaluate quality of SDR images when they are shown on displays with different brightness. Before passing through the PU-transform an \ac{SDR} image is first transformed to luminance emitted from a display, assuming a model of that display, such as the one in Equation \ref{eq:disp_model_SDR}. The full pipeline of making SDR image quality metrics photometric is given in Figure \ref{fig:extending_sdr_to_hdr}.

Authors in \cite{jia2017_2} used the PU-transform to adapt a no-reference deep \ac{SDR}~\ac{IQA} metric to operate on \ac{HDR} images. The model in \cite{jia2017_2} had to rely on a metric trained on SDR images due to the absence of a sufficiently large HDR dataset. In this work, we provide such a HDR dataset enabling us to train a deep HDR~IQA metric on both SDR and HDR images.

\section{Psychometric Scaling}\label{sec:rescaling}
In this section we extend a probabilistic model from our previous work \cite{2019TIP} to combine quality measurements from different datasets and from two different experimental protocols: ranking and rating. In the previous work, we demonstrated how mixing the scores from these protocols could lead to higher precision for a single dataset \cite{2019TIP}. In this work, we extend the idea from our previous work to mix the scores from different datasets and realign them to a common unified scale. The scaling method and the experimental protocol are generalisable to mixing other subjective assessment datasets where rating and ranking preference aggregation methodologies are used.

\subsection{Observer Model}
To build a quality scale, certain assumptions need to be made about how observers respond and perceive quality. Such assumptions are encapsulated in the observer model. This model is needed because observers vary in their notions of quality (inter-observer variance), and their opinions are also likely to change when they repeat the same experiment (intra-observer variance).

Let $\mathbf{q}$ be an N-dimensional variable whose individual components $q_i \in \mathbb{R}$ represent the underlying true quality of condition $i$, and $N$ is the number of conditions. We use the term condition to refer to an object or an item to be compared; in our case a condition represents an image content with certain distortion type and at a certain level.

According to the widely used Thurstone's model Case V \cite{Thurstone1927-THUALO-2}, perceived quality is normally distributed and the standard deviation $\sigma$ is known. 
 We can then introduce $r_i$ as the random variable associated with the measured quality, $r_i \sim \mathcal{N}(\hat{q}_i , \sigma)$.  Here the mean of the distribution approximates the true quality $q_i$ and the standard deviation is assumed to be the same for all conditions.

\subsection{Scaling Fundamentals}
Psychometric scaling aggregates pairwise comparisons into a (generally) uni-dimensional quality scale. The collected data  can be represented in a count matrix $\mathbf{C}$, where element $c_{ij}$ contains the number of times condition $i$ was chosen over condition $j$. Then, the probabilities of selecting one condition over another can be directly estimated from this matrix: 
\begin{equation}
    \hat{p}_{ij}= \frac{c_{ij}}{c_{ij}+c_{ji}}, i \neq j\,,
\end{equation}
where $\hat{p}_{ij}$ represents the probability that condition $i$ is perceived as of better quality than $j$.

 The main aim of psychometric scaling model is then to recover the true quality scores from these estimated probabilities. For the assumed model, the probability of choosing $i$ over $j$ should match the cumulative normal distribution $\Phi$ over the difference $r_i - r_j$:
\begin{equation}\label{eq:prob_pref}
    P(r_i > r_j) =  P(r_i - r_j > 0) = \Phi \left ( \frac{q_i - q_j}{\sqrt{2}\sigma}\right)\,,
\end{equation}
where $\sigma$ dictates the relationship between distances in the scale and probabilities of better quality. A common approach is to select $\sigma$ so that a
probability of 0.75 (in the midway between a random guess and being completely certain) is mapped to a distance of 1 unit in the quality scale. The difference of 2 units then corresponds to the probability of 0.91 and
so on.

A common approach for scaling pairwise comparisons is to use maximum likelihood estimation. That is, we estimate the difference in quality scores maximizing the probability of observing collected data $\mathbf{C}$. 
The probability that $i$ was selected over $j$ in exactly
$c_{ij}$ trials from the total number of $n_{ij} = n_{ji} = c_{ij} + c_{ji}$ trials is given by the Binomial distribution:
\begin{eqnarray}\label{eq:binomial}
     & L( q_i, q_j| c_{ij}, c_{ji}) = \nonumber
     \\& = {c_{ij} + c_{ji} \choose c_{ij}} P(r_i > r_j)^{c_{ij}} \left (1 - P(r_i > r_j) \right )^{c_{ji}}.
\end{eqnarray}
To scale all compared conditions, we maximise the product of the likelihood for all pairs of conditions:
\begin{equation}
   \argmax_{{q}_2,\ldots,{q}_n} \prod_{i,j{\in}\Omega} L({q}_i - {q}_j|c_{ij},n_{ij}), 
\label{eq:mle}
\end{equation} 
where $\Omega$ is the set of all pairs for which at least one comparison has been made. A more detailed discussion of the psychometric scaling can be found in \cite{perez2017}. 

\subsection{Mixing Pairwise Comparisons and Rating}
Our main assumption for mixing the two protocols, rating and ranking, is that different psychophysical experiments aim to recover the same latent ground truth variable $\mathbf{q}$ by different means. For pairwise comparisons, quality scores are recovered by psychometric scaling and for rating, approximated by mean opinion scores (MOS).
We further assume that the relationship between the MOS values (denoted as $\pi$) and the quality values resulting from psychometric scaling (denoted as $r$) is monotonic. We validated several polynomial relationships in supplementary.
Similar to previous literature \cite{Zerman2017_allignment} we found that for image quality assessment the relationship was well described by a linear model:
\begin{equation}\label{eq:relationship}
    r_i = a \cdot \pi_i + b\,,
\end{equation}
where $a$ and $b$ are the parameters and $\pi_i \sim \mathcal{N}(q_i, c \cdot \sigma)$ with $c$ compensating for varying measurement accuracy/observer models for both of these experimental protocols. Because rating protocols often use different rating scales (1--5 or 0--100) and because different protocols result in different inter- and intra-observer variation, we optimize for separate set of parameters $a, b$ and $c$ for each dataset.

Given that MOS values are generally measured in a continuous scale and we are assuming the same randomly distributed observer model, the probability of observing $m_{ik}$ (single rating measurement for $i^{th}$ condition and $k^{th}$ observer) can be expressed using the density function of the normal distribution: 
\begin{equation}
 f(m_{ik} | q_i, a, b,c) = \frac{1}{\sqrt{2 \pi a^2 c^2 \sigma^2}} e^{-\frac{((a \cdot m_{ik} + b)-q_i)^2}{2a^2c^2\sigma^2}}.
\end{equation}
Assuming independence between observers, the likelihood of observing ratings from $J$ observers for $N$ conditions aggregated in $\mathbf{M}$ is:
\begin{equation}\label{os_prob}
 P(\mathbf{M} | \mathbf{q}, \sigma, a, b,c) = \prod_{i=1}^N \prod_{k=1}^J f(m_{ik} | q_i, \sigma, a, b,c).
\end{equation}

To recover latent scores $\mathbf{q}$ from both measurements, the posterior probability can be factorised as: 
\begin{eqnarray}\label{eq:MLEmixing} \nonumber
    P(\mathbf{q},a,b,c|\mathbf{C}, \mathbf{M}, \sigma) \propto \\ P(\mathbf{q}) \cdot P(\mathbf{C} | \mathbf{q}, \sigma) \cdot P( \mathbf{M} | \mathbf{q}, a, b, c, \sigma ),
\end{eqnarray}
where 
$P(\mathbf{q} = \mathcal{N}(\frac{1}{N}\sum_{i=1}^{N}q_i, \sigma))$ is a Gaussian prior. {As we found in our earlier work \cite{perez2017}, the prior improves the precision of the scaling when the number of measured comparisons (per pair of conditions) is low. The prior introduces bounds on the score values in the presence of unanimous answers, which put no upper bound on the distance between the recovered scores.} The conditional independence of $\mathbf{C}$ and $\mathbf{M}$ given $\mathbf{q}$ is assumed. $P(\mathbf{C} | \mathbf{q}, \sigma)$ is computed as in Eq. \ref{eq:binomial} assuming independence between measurements. $P(\mathbf{M} | \mathbf{q}, a, b, c, \sigma)$ is computed using the density function of the normal distribution. We infer the quality scores $\mathbf{q}$, and the parameters  $a,b,c$ using maximum likelihood estimation. As scales are relative, we constrain the scores for all reference images (without any distortion) in all datasets to be zero ($q_i=0$ for each $i$ that represents a reference image). 

Since likelihood functions are scale invariant, we can fix $\sigma$ to any value without loss of generality. In our case we fix $\sigma=1.048$, so that a distance of 1 unit between two conditions indicates that 75\% of observers can see the difference between two conditions, allowing the interpretation of distances in the scale. 
These units are often referred to as Just-Objectionable-Differences (JODs) \cite{2019TIP}.

\section{Unified Photometric IQA dataset (UPIQ)}  \label{sec:LHDR_dataset}

Our goal is to create a large dataset consisting of both SDR and HDR images, with the image quality scores on a unified quality scale with JOD units. This is achieved by selecting existing SDR and HDR datasets, collecting additional cross and within-dataset comparisons, and scaling all the measurements together. We call our dataset UPIQ (``You Pick") --- Unified Photometric Image Quality. Before our work, the largest HDR IQA dataset contained only 240 conditions \cite{korshunov2015}. Our dataset, has 4159 images, making it the largest and the most diverse HDR image quality dataset to date. Unlike most IQA datasets, images in our dataset are provided in absolute photometric units cd/m$^2$ and scores are provided in interpretable JOD units.

\subsection{Selected Datasets}

\begin{table*}[!t]
    \small
    \caption{Characteristics of the chosen IQA datasets}
    \renewcommand{\arraystretch}{1.3}

    \centering
    \begin{tabular}{|c|c|c|c|c|c|c|}
        \hline
        Name  & \makecell{Dynamic \\ range}  & \makecell{Experiment} & \makecell{No. \\images}   & \makecell{No. \\ distortions}  & \makecell{No. \\contents} & \makecell{Image sizes\\(h$\times$w pixels)}  \\
        \hline
        LIVE \cite{Sheikh2006b} & SDR & MOS & 779 & 5  & 29 &  512$\times$768\\
        \hline
        TID2013 \cite{Ponomarenko2015}  & SDR & PWC & 3000 & 24 & 25 & 348$\times$512\\
        \hline
        Narwaria \cite{Narwaria2013} & HDR & MOS & 140 & 2 &  10 &  1080$\times$1920\\
        \hline
        Korshunov \cite{korshunov2015} & HDR & MOS & 240 & 3 & 20 &  1080$\times$944\\
        \hline
    \end{tabular}
    
    \label{datasets}
\end{table*}

\begin{figure}[t]
\centering
\includegraphics[width=\linewidth]{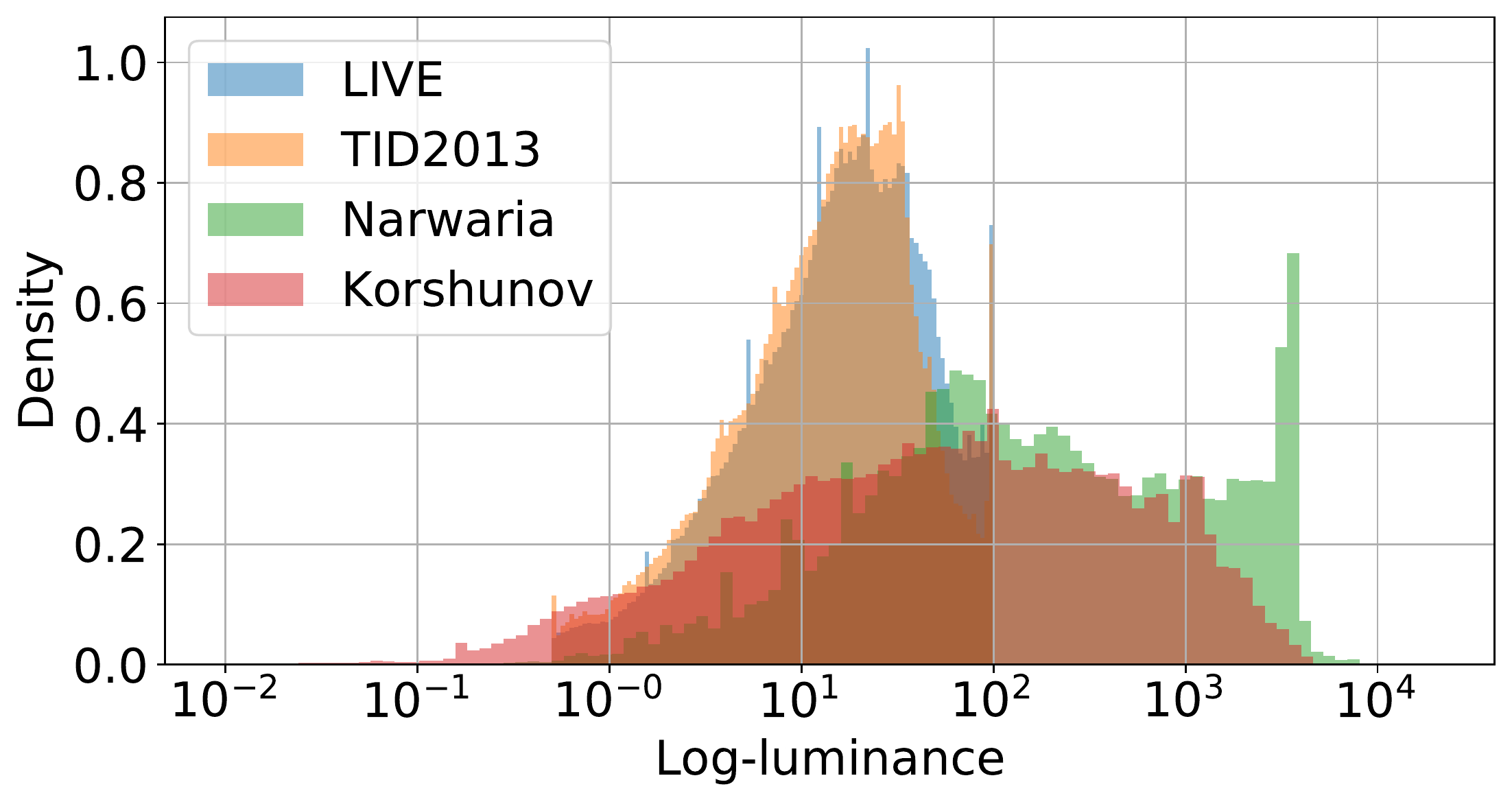}
    \caption{Distribution of log-luminance per dataset.}\label{fig:log_luminance}
\end{figure}

We selected four existing datasets---two SDR (TID2013 \cite{Ponomarenko2015} and LIVE \cite{Sheikh2006b}) and two HDR (Korshunov \cite{korshunov2015} and Narwaria \cite {Narwaria2013}), which we summarize in Table~\ref{datasets}. All four datasets span very large dynamic range, as shown in Figure~\ref{fig:log_luminance}. Despite a large number of available IQA datasets, only a few of them meet our criteria and could be included in UPIQ. Some datasets were constructed for the purpose of no-reference quality assessment \cite{8301594,8666733,8241862} and do not contain reference images \cite{7327186}. Other datasets contained a single distortion per content, thus they provided no means to scale the magnitude of a distortion \cite{7327186,6489321,8654007}. For some datasets, the image size was too small for a proper judgement of image quality \cite{8063957}. While we attempted to scale some datasets, we found their quality scores to be too inconsistent with our measurements to be included in UPIQ \cite{Zerman2017_allignment}. {We provide additional details on the dataset selection in the supplementary.}

\subsection{Dataset Alignment Experiments}\label{sec:experiment}

\begin{figure*}[h!]
\centering
\includegraphics[width=\linewidth]{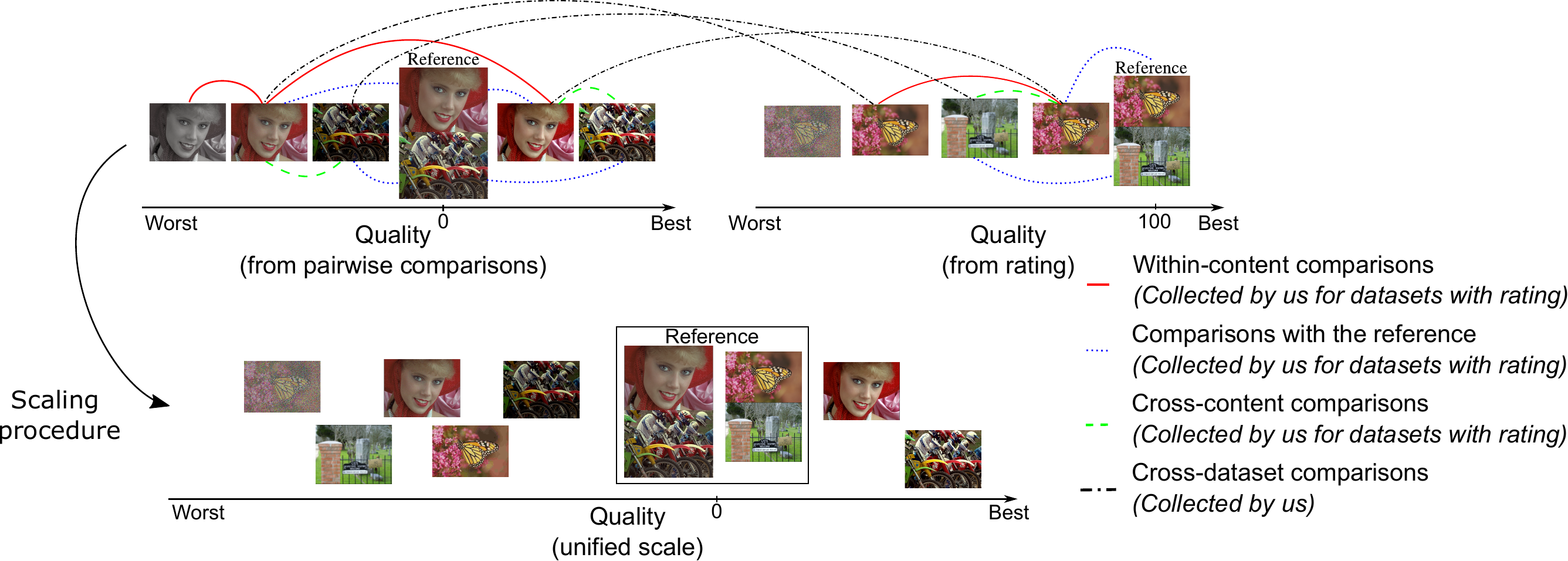}
    \caption{Types of comparisons necessary for dataset alignment. The lines link pairs of images selected for pairwise comparisons. Within-content comparisons (red solid lines) are most commonly used in pairwise comparison experiments. However, such datasets often lack comparisons with reference (blue dotted lines), which are useful to provide an absolute anchor of quality. Cross-content comparisons (green dashed lines) are less common, but can substantially improved the quality of the scale \cite{Zerman2018}. Finally, cross dataset comparisons (black dash-dotted lines) are necessary to scale the datasets together.
    }
    \label{fig:comparison_types}
\end{figure*}

To align quality scores from different datasets, we need to perform several types of pairwise comparisons, illustarted in Figure~\ref{fig:comparison_types}. Comparisons within a single dataset (within-content, cross-content and with-reference) are needed to bring the quality values to a common scale of JOD units. This is especially important for the datasets with only MOS (rating) values as these are provided in an arbitrary scale. We need to find the relationship between MOS and JOD values by estimating the associated parameters ($a, b$ and $c$ in Eq. \eqref{eq:relationship}). The cross-dataset comparisons are necessary to ensure that the quality values are comparable across the datasets. Because different datasets usually do not share the same content, cross-dataset comparisons tend to be also cross-content comparisons. Cross-content comparisons have been shown to be of the similar difficulty as within-content comparisons \cite{2019TIP} and they significantly improve the accuracy of a quality scale \cite{Zerman2018,Mikhailiuk2018}.

\paragraph{Displays and stimuli}
The data necessary for alignment were collected on two different displays. Comparison of SDR to SDR images were performed on a color calibrated 32" SDR Samsung S32D850T display with $2560\times1440$ pixels, 300\cdms typical peak luminance and a black level of $\sim$0.3\cdms. The comparisons involving HDR images were presented on a custom-built, color-calibrated 10" HDR display with $2048\times1536$ pixels, 15,000\cdms peak luminance and a black level below 0.01\cdms \cite{Wuerger2020}. 

We used the display model from Equation \ref{eq:disp_model_SDR} to transform gamma-corrected sRGB colors to linear RGB values shown on the HDR display. Because we had no information on the displays used in the SDR image quality studies, we used the typical parameters of an SDR display: $\gamma=2.2$, the peak luminance, $L_{\mathrm{peak}} = 100$\cdms, and the black level, $L_{\mathrm{black}} = 0.5$\cdms. For HDR images, we reproduced the absolute luminance values used in the original studies. {The viewing distance was 90\,cm for both the SDR display (51 pixels per degree) and the HDR display. Both HDR and SDR images, shown on the HDR display, were upscaled by a factor of 3.2 (50 pixels per degree), making the measurements taken for the original SDR and HDR studies comparable with ours.}

\paragraph{Experimental procedure and participants}
The observers were presented with pairs of images and were asked to select the image of better quality with respect to the reference. Observers could press and hold the space-bar to view the reference images. {When the image size exceeded the size of our display, we provided a simple panning interface in which observers could use a trackball to inspect different portion of the image.} Each participant viewed images in different order. Each selected pair of images was compared by 6 participants, with each participant completing approximately 300 trials. Overall 6000 new comparisons were collected from 20 participants. Note that this required relatively moderate experimental effort as compared to collecting the data from scratch (3000 images in the TID2013 dataset required over 500,000 comparisons).
To improve the information gain of the collected data and to exclude obvious comparisons \cite{ye2014active}, paired images were of similar quality. 

{We conduct two types of experiments. The first is cross-dataset comparisons. This type of comparisons was only necessary for rating-based datasets, which means we excluded TID2013 from this experiment since we used previously collected pairwise comparisons and rating measurements \cite{Mikhailiuk2018,2019TIP}. We ensured that all three types of necessary comparisons were covered: to reference, within-content and cross-content. After the first experiment, all the data could be scaled, since we had comparisons to a common reference for all datasets. }

{For the second experiment we compared conditions exclusively from different datasets, connecting each dataset to the rest. Images were chosen to uniformly cover the whole quality scale. We performed several iterations of the pair selection. After conducting an experiment on a small batch of comparisons, we re-scaled the dataset with newly collected comparisons and selected the next batch from the new scale.}

\subsection{UPIQ Dataset Scaling}

We combined the newly-collected comparisons with the original data from the four datasets and from the two follow-up studies on TID2013 \cite{Mikhailiuk2018} and LIVE \cite{ye2014active}. In total, the combined dataset consists of 571,215 individual pairwise comparisons and 27,676 rating measurements, which were passed to the scaling procedure from Section~\ref{sec:rescaling}. 

\begin{figure*}[t]
\centering
\includegraphics[width=\linewidth]{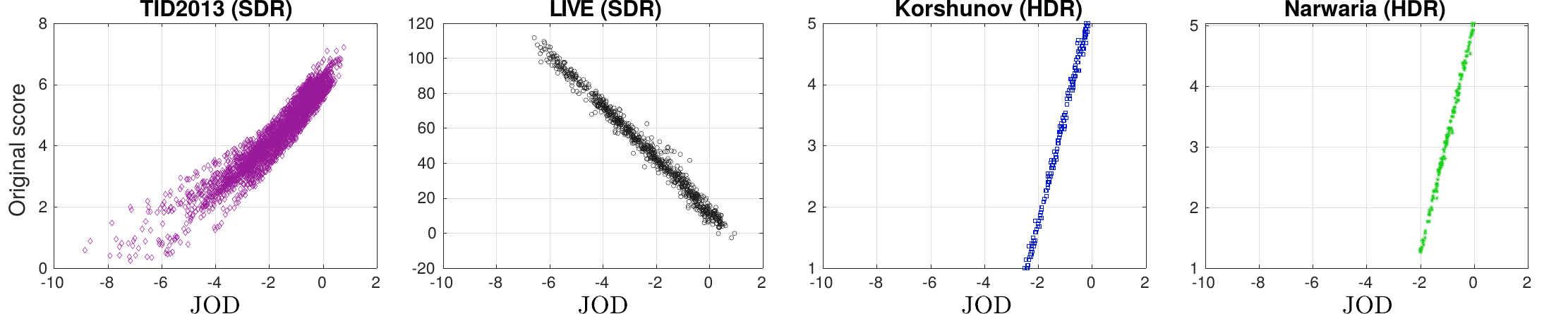}
    \caption{Original quality scores versus the results of our joint dataset scaling in JODs (UPIQ dataset). Note that the original scores of the TID2013 are the most different from the new JOD scores. This is because the original dataset used a simplified scaling procedure and lacked cross-content comparisons \cite{Mikhailiuk2018}. 
    }\label{fig:original_vs_scaling}
\end{figure*}

Figure \ref{fig:original_vs_scaling} shows the relationships between original quality values of each dataset and the new JOD values from our unified dataset. The plots show substantial differences between the original and rescaled quality scores, suggesting that cross-dataset scaling and additional measurements have further improved the quality estimates. 
Note that the original scores of the TID2013 dataset were obtained with vote counts, reliant only on within-content comparisons. This approach has proven to be less accurate as compared to psychometric scaling \cite{Mikhailiuk2018}.

\subsection{UPIQ Dataset Validation}\label{sec:validation_of_UPIQ_dataset}

The qualitative comparison, showing images at a constant JOD level, is shown in supplementary. 
The figures demonstrate that images at the same JOD level contain comparable perceived magnitude of distortions. In the following subsections we compare our scaling with the metrics-based dataset alignment, and then demonstrate the improvement in pairwise accuracy. 

\subsubsection{Comparison to previous re-scaling work}
\begin{figure}[t]
\begin{center}
   \includegraphics[width=.8\linewidth]{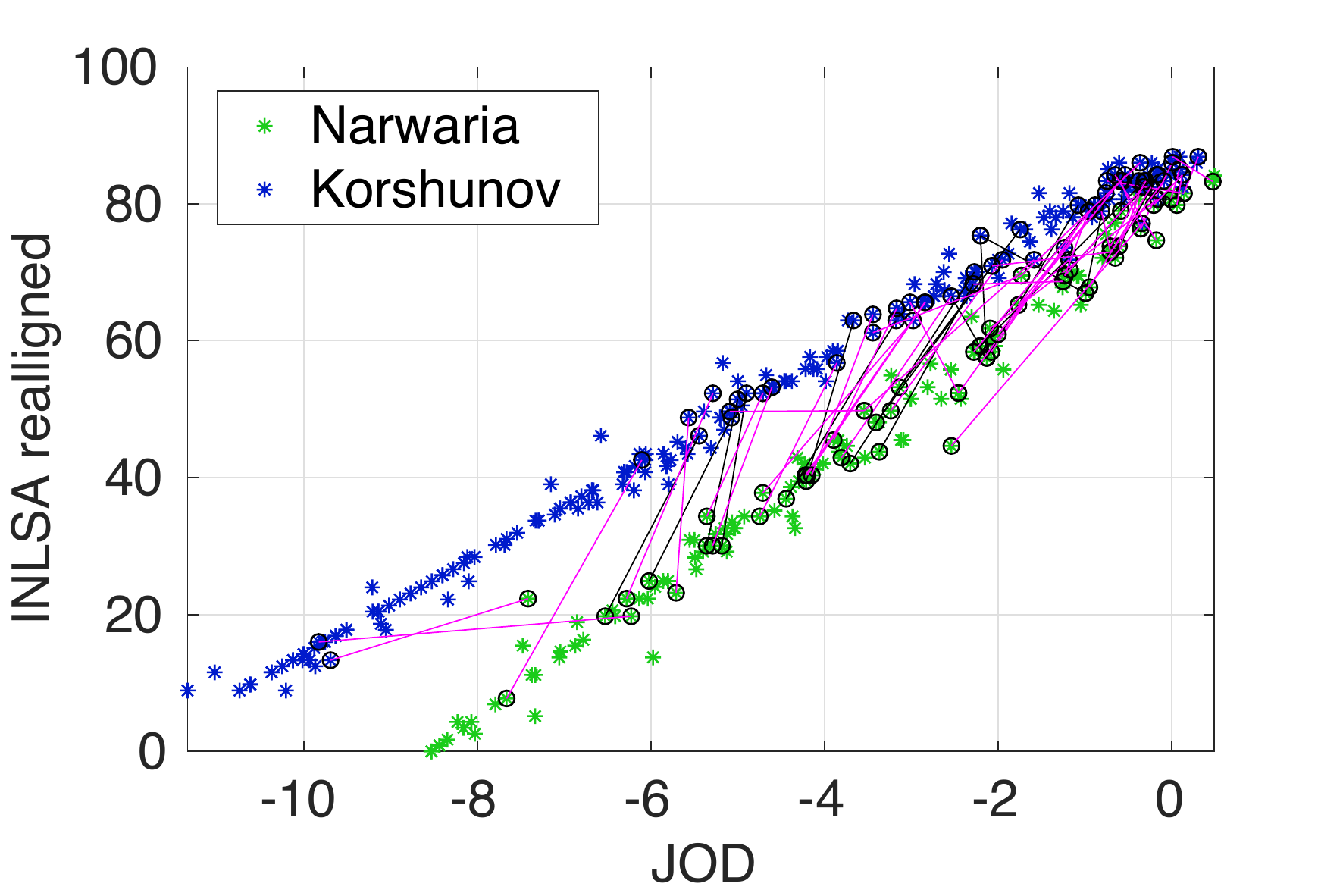}
\end{center}
   \caption{ Scatter plot of the two considered quality scales for the HDR datasets. The plot also shows an example of the data used for one of the cross-validation folds. Purple lines represent the training comparisons and black lines the test comparisons}
\label{fig:scale_validation_INLSA}
\end{figure}

Multiple \ac{IQA} datasets can be merged together using an iterated nested least-squares (INLS) algorithm \cite{INLSA}. The algorithm uses existing objective quality metrics to find the relationship between conditions in different datasets. The assumption made is that a weighted combination of metrics should have high correlation with human judgments.
The algorithm iteratively finds weights for the combination of objective quality metrics and aligns subjective quality scores from each of the datasets until convergence. 
Since no metric exists that has been exhaustively tested on both SDR and HDR images, we validate the results using two HDR datasets (Korshunov and Narwaria), which were aligned with INLS in the previous work \cite{Zerman2017_allignment}. Figure~\ref{fig:scale_validation_INLSA} shows that our scaling procedure and the one from \cite{Zerman2017_allignment} lead to substantially different scores. 
To determine which alignment is more consistent with the subjective judgements, we compute the rank-order correlation between the unprocessed human subjective measurements and scaled values. Since the collected human judgment data comes in the form of pairwise comparisons, we compute the correlation between empirical probability of selecting one condition over another and differences in quality scores. 

\begin{figure*}[t]
\centering
\begin{center}
\includegraphics[width=.75\linewidth]{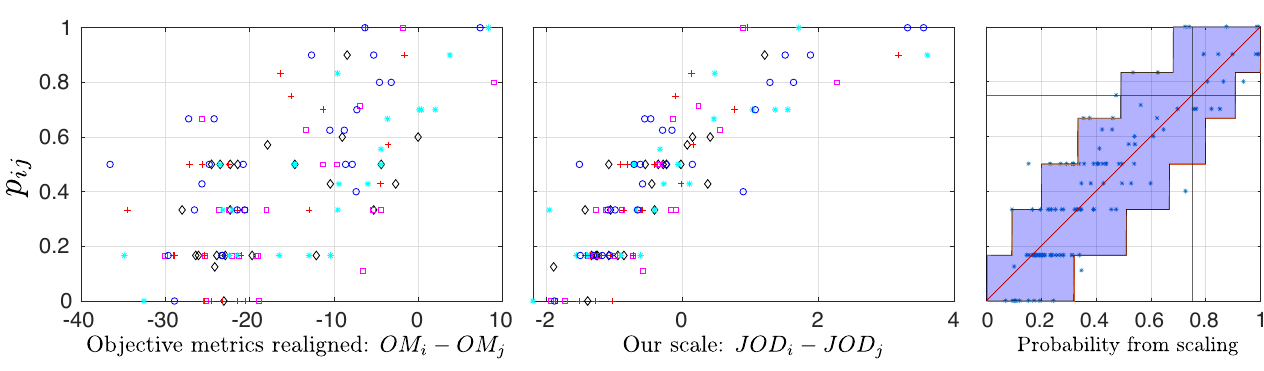}
\end{center}

    \caption{{Right two plots: scatter plots for five folds (distinguished by colors and shape); difference in quality scores in the scale constructed with objective quality metrics \cite{Zerman2017_allignment} $\mathrm{OM}_i$ - $\mathrm{OM}_j$ versus empirical probability, $p_{ij}$, of one image $i$ being selected over image $j$ (left); and difference in quality scores in our scale $\mathrm{JOD}_i - \mathrm{JOD}_j$ versus $p_{ij}$ (center). Our scale is clearly better aligned with the measurements. The right plot: the same as the centre plot but for the entire dataset (no folds) and plotted as the probability instead of JOD difference. The shaded area denotes a 90\% confidence interval for the measurements ($p_{ij}$). The 75\% preference (1 JOD) is denoted by the black vertical and horizontal lines. Our scaling procedure brings the quality differences within the confidence interval of the measurements.} }
\label{fig:75_ci_prob_vs_jod}
\end{figure*}

The method proposed in \cite{INLSA} relies exclusively on rating and objective metrics to re-align datasets. Whereas our approach, uses psychometric scaling, which requires additional pairwise comparisons to build the unified quality scale. Thus, to ensure a fair comparison, we perform a five-fold cross-validation on the collected cross-dataset comparisons. We split cross-dataset comparisons into five equal-sized partitions. In each fold of the cross-validation we scale the data from four partitions and use the fifth partition for validation. The cross-validation results are given in Table~\ref{tb:ffcv}. For each fold, our model correlates better with the subjective judgments, with mean \ac{SROCC} of 0.71 versus 0.56 for the method from \cite{Zerman2017_allignment}. It should be noted that the correlation values computed in this manner cannot reach high values because of the measurement noise in the pairwise comparison data. Figure~\ref{fig:75_ci_prob_vs_jod} also shows that the relationship is closer to the expected cumulative normal function for our method. {This is further confirmed when JOD differences are converted into probabilities (Equation~\ref{eq:prob_pref}) and plotted in the right panel of Figure~\ref{fig:75_ci_prob_vs_jod}. The scaled probabilities are within the confidence interval of the measured probabilities, confirming that the scaling procedure leads to the quality values that well reflect the empirical quality differences.}

\begin{table}[!t]
    \small
    \renewcommand{\arraystretch}{1.3}
    \caption{\ac{SROCC} between scaled quality scores and empirical probabilities, for our and metric-based scaling. The values are reported for each fold of the cross-validation.}
    \centering
    \begin{tabular}{|c|c|c|c|c|c|}
        \hline
        Validation Fold  & 1 & 2 & 3 & 4 & 5 \\
        \hline
        Psychometric scaling (our)& 0.77 & 0.72 & 0.62 & 0.74 & 0.71 \\
        \hline 
        Objective-metric-based \cite{Zerman2017_allignment}&0.67 & 0.60 & 0.52 & 0.52 & 0.53 \\
        \hline
    \end{tabular}

    \label{tb:ffcv}
\end{table}

\subsubsection{Measuring pairwise accuracy}
In this section we provide interpretation of the SROCC results from our validation experiment. We will demonstrate that the scale correctly ranks 97\% of the pairs that are at least 1\,JOD apart.

We first transform the collected data and the produced scale into pairwise rankings. This is, if the quality of $i$ is higher than that of $j$ (as measured in the collected pairwise comparison matrix $\mathbf{C}$) then we set the binary target label $t_{ij}$ to +1, otherwise we set the target $t_{ij}$ to -1. This represents the ground truth pairwise rank averaged across the population. We then compare this ground truth binary label to our predicted binary labels $\hat{t}_{ij}$, following the same procedure but using the output of the scaling algorithm instead of probabilities. Having ground truth and predictions, we compute ranking accuracy. For this, we ran a 10-fold cross-validation. In each iteration we withheld 10\% of the compared cross-dataset pairs of conditions for validation. The remaining 90\% of compared pairs were for scaling. 
To compute the ranking accuracy, we assume the minimum threshold distance (in terms of JODs) that is required for a pair of conditions to be considered, then report the ratio of the number of correctly ranked considered pairs to the total number of considered pairs.

\begin{figure}[t]
\begin{center}
   \includegraphics[width=\linewidth]{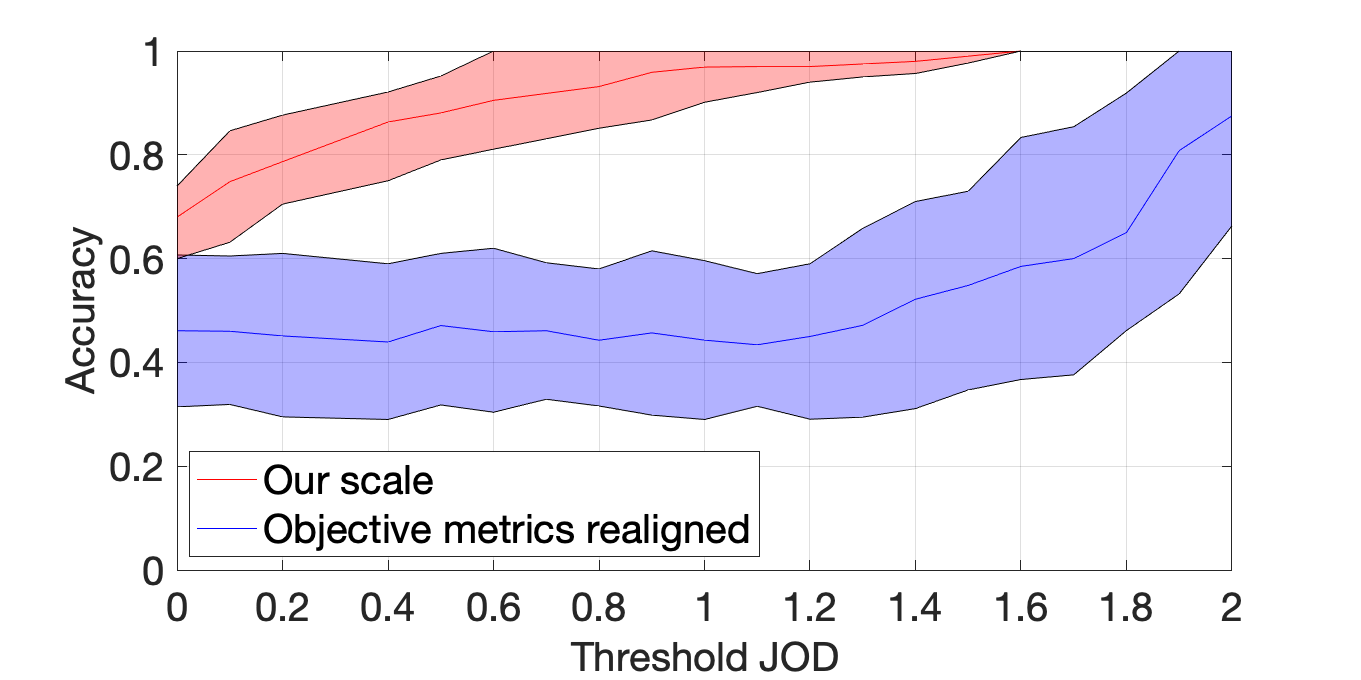}
\end{center}
   \caption{
Accuracy of classifying cross-dataset conditions into better/worse after alignment with the proposed method. Higher value of Threshold JODs means that more conditions are excluded from training and testing sets. Shaded region is 95\% confidence interval.}
\label{fig:additional_validation_srocc_improvement}
\end{figure}

Figure~\ref{fig:additional_validation_srocc_improvement}  shows the accuracy scores for different thresholds of reliable JOD differences, for both our scale and that of  \cite{INLSA}. For conditions $>0.75$\,JODs apart (where 63\% of observers agreed on the highest quality image, only 13\% more than random choice), our scale has 90\%  accuracy. That is, 90\% of the pairs which are more than 0.75\,JODs apart are correctly ranked by our psychometric  scaling. The difference with \cite{INLSA} is very significant with our scale being consistently much more accurate across different thresholds.

\subsubsection{Cross-dataset scaling versus multi-task learning}

The final hypothesis we tested is whether a multi-task deep-learning network can implicitly infer the relation between the datasets and thus can be used to merge datasets without the need for additional data. In this approach, the neural network is trained to predict the scores of the four individual datasets (tasks). The network consisted of two parts. The first part predicts a common score. The second part maps the common score via simple linear regressors, different for each dataset, to the original quality values collected in each dataset. Both parts were trianed jointly end-to-end. The details of the architecture are provided in supplementary. 
This experiment, however, was unsuccessful. The average \ac{SROCC} between the common score predicted by the multi-task network and the ground truth pairwise comparisons, was only 0.27 for  cross-dataset pairs.

\section{Applications} \label{sec:obj_metrics}
In this section we show how our UPIQ dataset is useful to train CNN-based metrics and benchmark existing HDR quality metrics. We further show how metrics trained on our dataset can be used for brightness-aware image compression. 

\subsection{Training Data-driven HDR Metric}
\label{sec:dpiqm}
UPIQ is a sufficiently large image quality dataset to enable us to train from scratch a CNN-based image quality metric to predict quality of both SDR and HDR images. The metric combines the ideas behind PU encoding \cite{Aydn2008} (Section~\ref{sec:PU-encoding}) and a recently proposed CNN architecture for image quality assessment (PieAPP), \cite{Prashnani_2018_CVPR}). We will refer to this metric as PU-PieAPP.

\paragraph{Architecture} 

\begin{figure}[t]
\begin{center}
   \includegraphics[width=\linewidth]{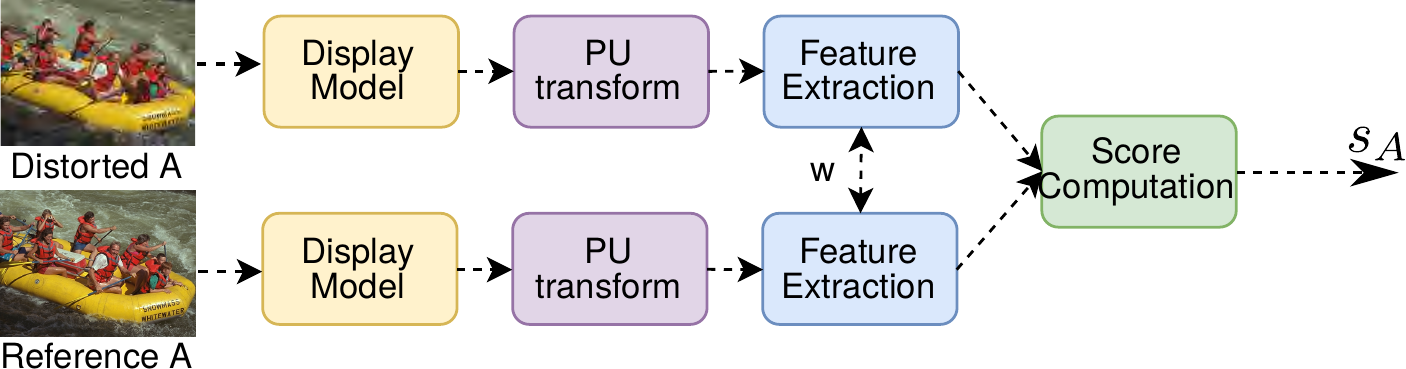}
\end{center}
   \caption{
   The pipeline used to train PU-PieAPP on absolute scores. Images first pass through the display model and are then fed to the PU-transform. The feature extraction network with shared weights (w) extract representations, which are passed to the score computation network. 
   }
\label{fig:architecture}
\end{figure}
The diagram of the deep metric architecture is shown in Figure~\ref{fig:architecture}. The metric takes as input a pair of test and reference images and produces a single quality score $s_A$ in JODs. To account for the dynamic range of the displayed images, the input images need to be transformed into the display domain. This is achieved by a \emph{display model} from Equation~\ref{eq:disp_model_SDR} for SDR images, or by scaling color values according to the presentation conditions from the original papers for HDR images. Then, the resulting trichromatic color values (with Rec. 709 primaries \cite{BT709}) are converted into approximately perceptually uniform units using the PU-transform (Section~\ref{sec:PU-encoding}), which is applied individually to each color channel. Such encoded images are fed into the PieAPP architecture, which combines a pair of feature extraction networks with shared weights with the score computation network, both identical to the one used in \cite{Prashnani_2018_CVPR}. The detailed architecture is provided in supplementary. 

We train the network on $64 \times 64$ patches. To densely cover the whole image, the image is stratified by a uniform grid and patches are sampled at random positions in each grid cell (jittered sampling). The grid size is selected to give approximately square cells. In training, we extract 1024 patches per image. We found that 1024 was the largest number of patches that we could process on our GPU. When testing, we sampled twice the number of $64 \times 64$ patches needed to cover the image.
This number was optimal in terms of the time vs. performance trade off.

\paragraph{Training} In contrast to \cite{Prashnani_2018_CVPR}, we train the network as regression rather than learning-to-rank. Our scaling procedure achieves the same goals as learning-to-rank, but offers a more accurate observer model and allows us to split the problem into two separate steps of scaling and learning. We train the network from scratch, using Adam optimizer on 4 NVIDIA P100 GPUs. Every tested architecture was run for 500 epochs and the model with the best performance on the validation set was saved (using 60-20-20 split into training, validation and test sets).

\subsection{Benchmark of HDR Quality Metrics}

\begin{figure*}[t]
\centering

    \includegraphics[width=0.98\linewidth]{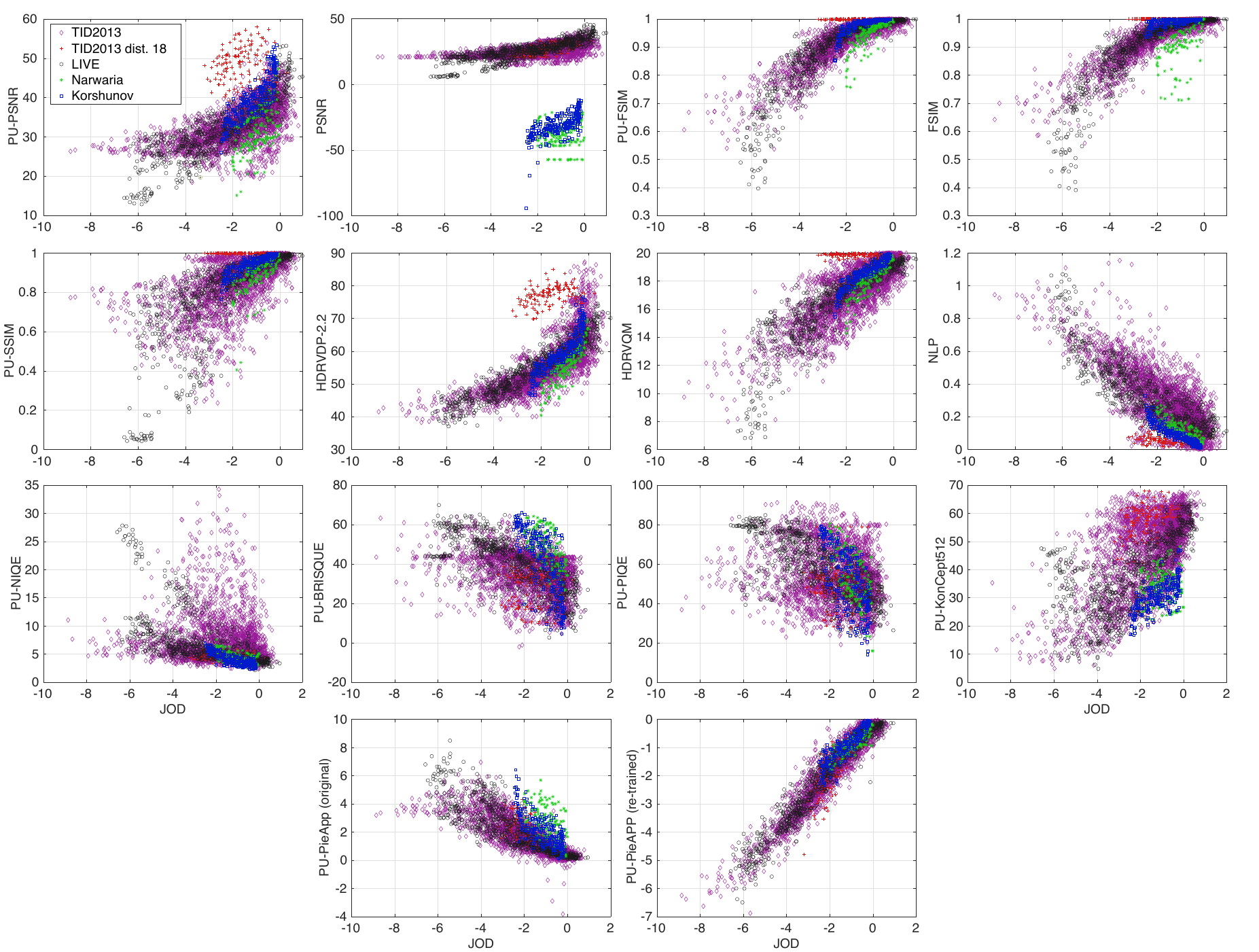}

    \caption{Objective metrics predictions vs. JOD quality values. We separately label distortion 18 from TID2013 (change of color saturation) as it introduces the biggest prediction error for the metrics that operate on luma/luminance values and ignore color information.
    } \label{fig:scaling_vs_objective_q}

\end{figure*}

Although HDR image quality metrics have been compared in many studies \cite{korshunov2015,Zerman2017_allignment,etpoefrqmohdrvc}, none of them could test the metrics on an extensive dataset such as UPIQ. Therefore, we use UPIQ to test existing HDR metrics. 

Here we consider full-reference metrics, which are either adapted to HDR content using PU-transform: PU-PSNR, PU-SSIM \cite{1284395}, PU-FSIM \cite{5705575}, or are designed to work with \ac{HDR} data: \ac{HDR-VQM} \cite{NARWARIA201546}, HDRVDP-2.2 \cite{Mantiuk:2011:HCV:2010324.1964935,Narwaria} and NLP \cite{Laparra_17}. We also evaluate no-reference metrics, adapting them to the HDR content with PU-transform: PU-BRISQUE \cite{6272356}, PU-PIQE \cite{7084843} and PU-NIQE \cite{Mittal2013MakingA}, due to their widespread use and competitive performance. \
Finally, we adapted existing \ac{SDR} \ac{CNN}-based metrics to \ac{HDR} content using the PU-transform: PU-KonCept512 \cite{Hosu_2020} (no-reference) and original PU-PieApp (original) \cite{Prashnani_2018_CVPR} (full-reference).
We did not re-train deep metrics on UPIQ, but used weights provided by the authors.
For comparison, we also include full reference PSNR and FSIM metrics, not adapted to the HDR content. 
\begin{figure*}[!ht]
\centering

    \includegraphics[width=0.95\linewidth]{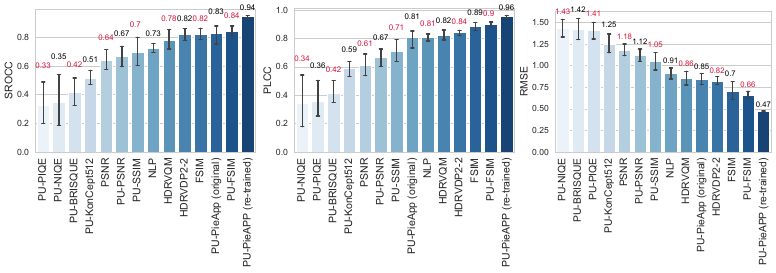}
    \caption{Cross-validation results for all trained metrics, expressed as \ac{SROCC}, \ac{PLCC} and \ac{RMSE}. Error bars denote 95\% confidence intervals. 
    }
    \label{fig:validation_objective_q}
\end{figure*}

\begin{table*}[!t]
    \small
        \caption{Test \ac{RMSE}, \ac{SROCC} and \ac{PLCC} for different data partitioning schemes and the best performing metrics.  (C-C -- cross-content, C-D -- cross-dataset, C-DR -- cross-dynamic-range). 
     We remove the listed test portion of the UPIQ from training, and test on it. Note that the first column (C-C) shows the same data as Figure~\ref{fig:validation_objective_q}.
    }
    \renewcommand{\arraystretch}{1}
    \begin{center}
    \begin{tabular}{| m{1.55cm}|m{1.4cm}|m{1.35cm}|m{1.4cm}|m{1.4cm}|m{1.4cm}|m{1.4cm}|m{1.4cm}|}
        \hline 
       Metric & \makecell{C-C\\ Test: \\sel. cont.} &
         \makecell{C-D\\ Test: \\
         TID2013} & \makecell{C-D \\ Test:\\
         LIVE} & \makecell{C-D \\ Test: \\
         Narwaria} & \makecell{C-D\\ Test: \\
         Korshunov} & \makecell{C-DR \\ Test: 
         \\HDR} & \makecell{C-DR \\ Test: \\
         SDR} \\

        \hline
        \hline
        \multicolumn{8}{|c|}{\textbf{RMSE}}  \\  \hline \hline
         PU-PieAPP & \textbf{0.47} & 0.92          &  {0.70}       & 0.68          & 0.62          & 0.72          & \textbf{1.29} \\ \hline
         PU-FSIM   & 0.66          & \textbf{0.65}&  \textbf{0.50} & 0.26          & 0.29          & 0.68          & 1.39          \\ \hline
         FSIM      & 0.70          & 0.65         &  0.51          & 0.45          & 0.52          & 1.17          & 1.61          \\ \hline
         HDRVDP    & 0.82          & 0.88         &  0.64          & 0.24          & 0.21          & 0.78          & 1.34          \\ \hline
         HDRVQM   & 0.86          & 1.04         &  0.68          & \textbf{0.23} & \textbf{0.20} & \textbf{0.39} & 1.43          \\ \hline

        \hline
        \hline
        \multicolumn{8}{|c|}{\textbf{SROCC}}    \\  \hline
        \hline
         PU-PieAPP & \textbf{0.94} & 0.78          & 0.87          & 0.82           & 0.79          & {0.74}         & 0.65   \\  \hline
         PU-FSIM   & 0.90          & \textbf{0.80} & \textbf{0.96} & 0.87           & 0.93          & 0.71           & 0.77   \\  \hline
         FSIM      & 0.89          & 0.80          & 0.96          & 0.54           & 0.52          & 0.45           & 0.54 \\  \hline
         HDRVDP    & 0.84          & 0.78          & 0.94          & 0.94           & 0.94          & 0.81           & \textbf{0.82}   \\  \hline
         HDRVQM   & 0.82          & 0.71          & 0.92          & \textbf{0.95}  & \textbf{0.95} & \textbf{0.87}  & 0.60 \\  

         \hline
        \hline
        \multicolumn{8}{|c|}{\textbf{PLCC}}    \\  \hline
        \hline
         PU-PieAPP & \textbf{0.96} & 0.78          & 0.89          & 0.78           & 0.75          & 0.73           & 0.67   \\  \hline
         PU-FSIM   & 0.90          & \textbf{0.89} & \textbf{0.96} & 0.87           & 0.90          & 0.66           & 0.77   \\  \hline
         FSIM      & 0.89          & 0.89          & 0.96          & 0.53           & 0.66          & 0.34           & 0.51 \\  \hline
         HDRVDP    & 0.84          & 0.83          & 0.93          & 0.89           & 0.95          & 0.72           & \textbf{0.78}   \\  \hline
         HDRVQM    & 0.82          & 0.78          & 0.92          & \textbf{0.89}  & \textbf{0.95} & \textbf{0.86}  & 0.62 \\  \hline

    \end{tabular}
    \end{center}
    \label{tb:data-partitioning}
\end{table*}

Most objective metrics predict values that are non-linearly related to absolute quality in JOD units. The scatter plot of the considered metrics predictions versus those of the JOD's is provided in Figure \ref{fig:scaling_vs_objective_q}. Since our goal is to predict the absolute quality, we need to map metric predictions to JODs. We follow a standard approach \cite{Sheikh2006b} and fit a logistic function mapping objective quality $o$ into absolute JOD units $q(o)$:
\begin{equation}
    q(o) = \frac{a_1}{1+e^{a_2(o-a_3)}}+a_4o+a_5,
\end{equation}
where $a_1, \dots, a_5$ are fitted parameters. Fitting a logistic function is necessary for computing performance measures, RMSE and PLCC, but it also helps to scale objective metric results into interpretable and comparable units of JODs. For example, while the result of PU-SSIM of 0.98 is difficult to interpret, the result of -1\,JODs tell us that 75\% of the population will notice the difference. 

For fair comparison, we use the same 5-fold split into 80-20\% training and testing dataset when fitting psychometric function for the tested metrics. In each fold a different portion of the entire dataset is tested while ensuring that no content is shared between training and testing sets. We also make sure that each subset (TID2013, LIVE, Narwaria and Korshunov) was split in the same 80-20 ratio. Note that since PU-PieAPP (re-trained) is trained on the quality scores from UPIQ dataset, we do not need to fit the logistic function into its prediction.

\paragraph{Cross-content validation} The most common approach to the validation of learning-based quality metrics is the split into training and testing sets that contain different content but share distortion types. Note that we took extra care to isolate the same content in LIVE and TID datasets as those share some of the reference images. The results for the 5-fold cross-validation on such cross-content splits, shown in Figure~\ref{fig:validation_objective_q}, indicate that PU-PieAPP (re-trained) outperforms existing hand-crafted metrics. PU-PieAPP shows 30\% improvement to the second-best performing metric, PU-FSIM, which is followed by FSIM without the PU-transform. We later show that the performance difference between PU-metrics and original metrics is much higher when tested on HDR datasets (UPIQ is dominated by images from SDR datasets). 
No-reference metrics, based on hand-crafted features, exhibit the worst performance --- the PU-transformation applied to the images distort the statistics that these metrics rely on. Deep learning based no-reference metric PU-KonCept512 does not perform well either. 

Original PieApp adapted to our dataset with PU-transform, performs reasonably well on SDR images (SROCC: 0.8764). However, exhibits poor performance on both HDR datasets (SROCC: 0.5791). This is expected, as the metric was trained on SDR images, and the range of PU-transformed HDR images is larger than that of SDR.

\begin{figure*}[!ht]
\centering
    \begin{subfigure}[b]{0.3\textwidth}            
            \includegraphics[width=\linewidth]{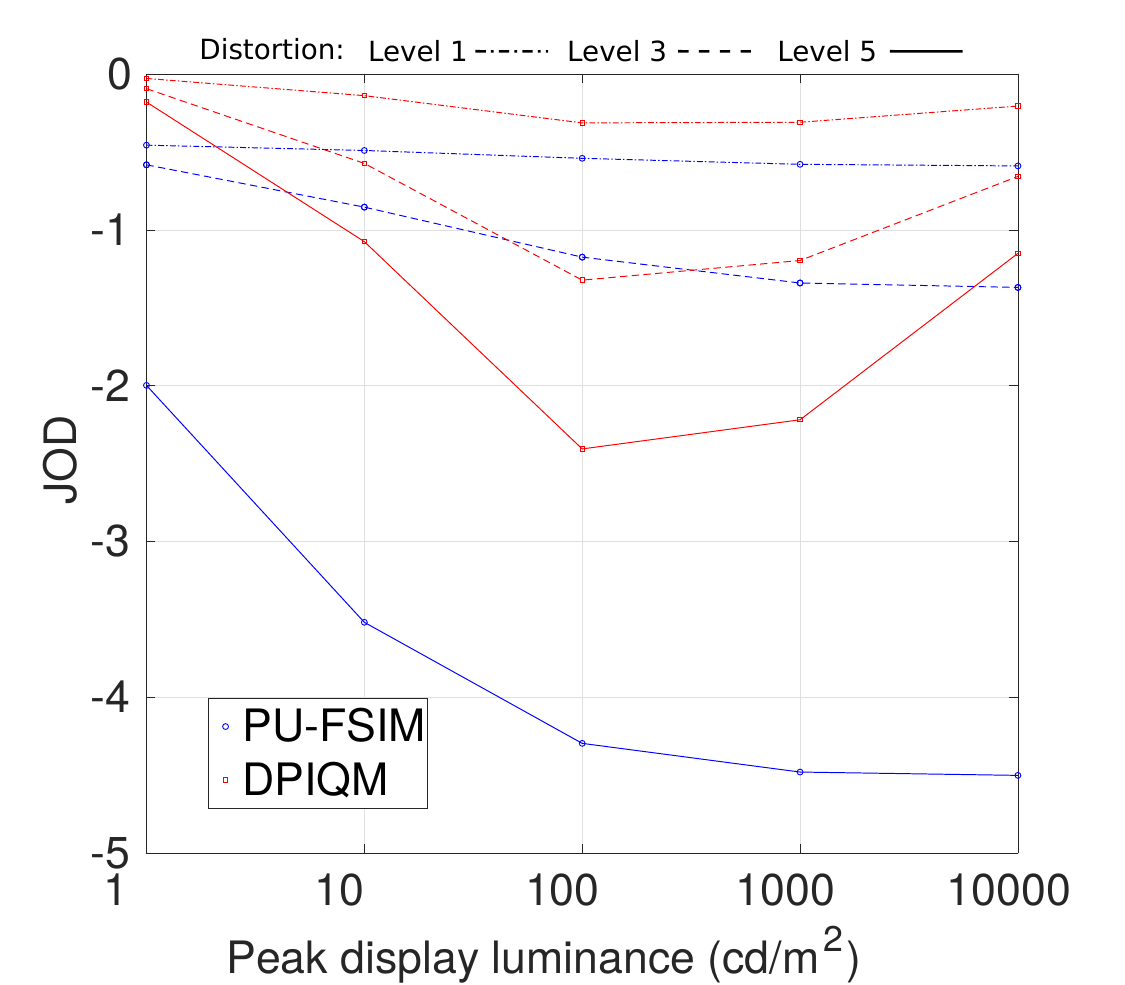}
            \caption{Prediction across the luminance range }
            \label{fig:pred_across_lum}
    \end{subfigure}%
    ~
        \begin{subfigure}[b]{0.3\textwidth}
            \centering
            \includegraphics[width=\linewidth]{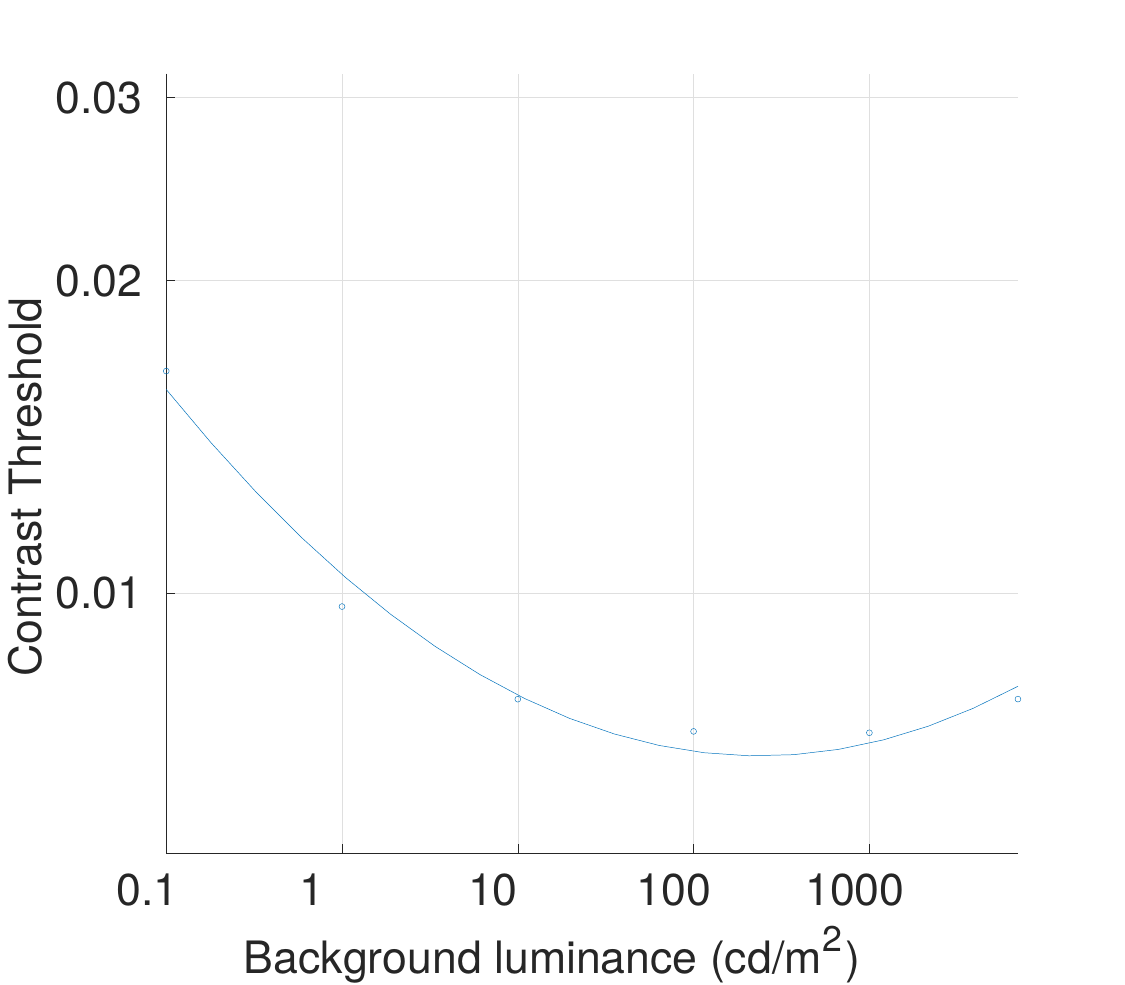}
            \caption{Contrast thresholds}
            \label{fig:csf_mj}
    
    \end{subfigure}
    \begin{subfigure}[b]{0.35\textwidth}
            \centering
            \includegraphics[width=\linewidth]{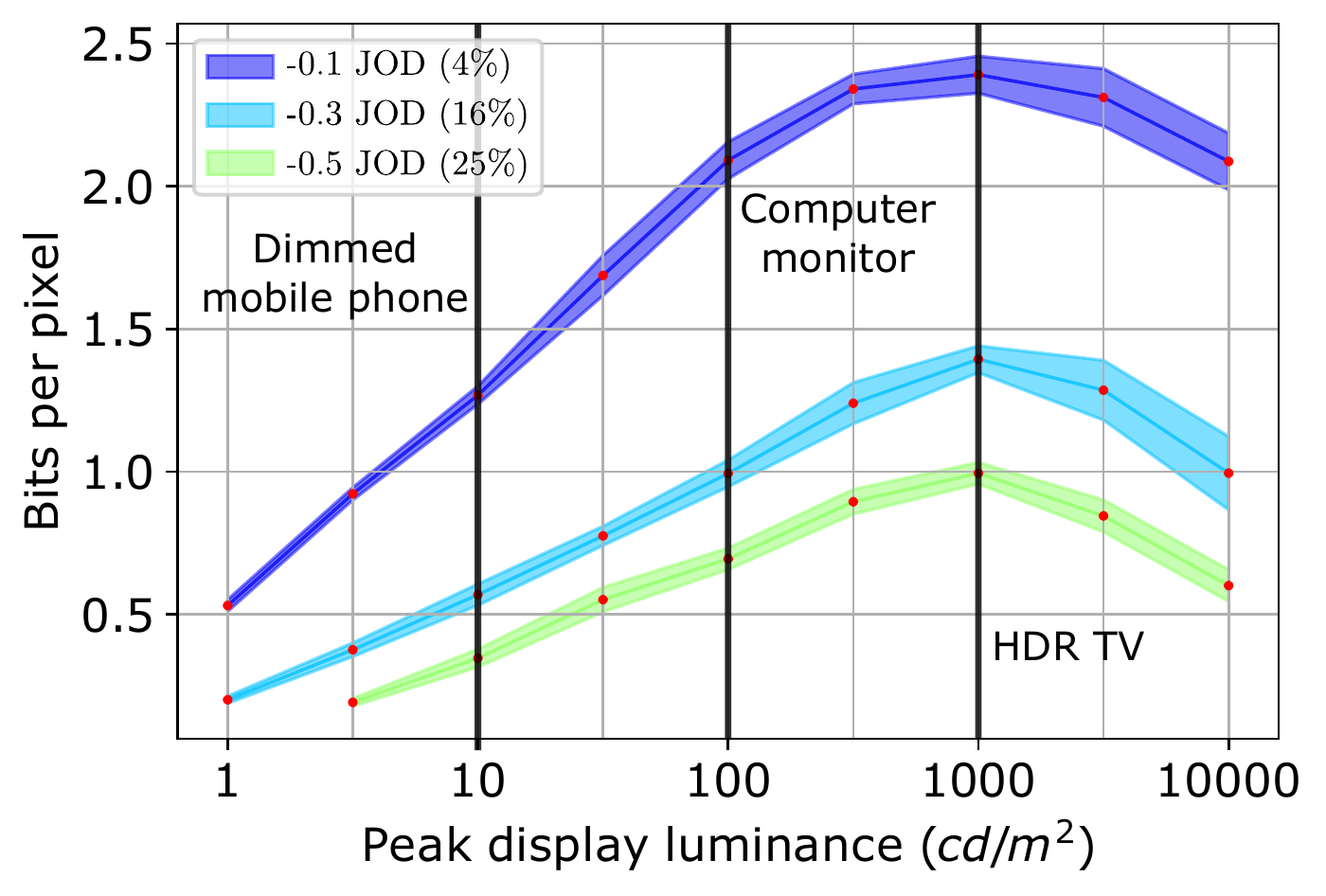}
            \caption{Brightness-Adaptive Coding}
            \label{fig:application}
    
    \end{subfigure}
    \caption{(a) The quality predictions as the function of display peak luminance. The predictions are shown separately for three distortion levels of JPEG and averaged across contents from TID2013 dataset. (b) Contrast threshold function for a varied display brightness; (c) Bits per pixel for JPEG compression to achieve constant perceived quality at different luminance levels. Different colors represent different quality levels. Shaded regions are 75\% confidence intervals.}

\end{figure*}
\paragraph{Cross-validation schemes} To understand what mixture of data is required to robustly train quality metrics, we experiment with different data partitioning schemes. For this experiment, we selected 5 best performing metrics from Figure~\ref{fig:validation_objective_q}. 
Table~\ref{tb:data-partitioning} lists the training and test data combinations we tested and the corresponding results. 

PU-PieAPP generalizes well when trained cross-content (C-C), i.e.~ the training and test set overlap in distortion types but not in content. However, the performance of this deep-learning metric drops significantly if one or more datasets are missing from the training set. This, and the poor performance of no-reference metrics in Figure~\ref{fig:validation_objective_q}, show that learning-based metrics are prone to overfitting when the training dataset is not sufficiently large.

As expected, SDR metrics exhibit better performance when tested on SDR datasets. The same holds for metrics aimed for HDR content -- they perform better on HDR datasets. PU-FSIM and FSIM have similar performance when tested on SDR datasets. However, when tested on HDR, PU-FSIM performs significantly better compared to FSIM, clearly demonstrating the need for the PU-transform.

\subsection{Maximum differentiation competition}
\begin{figure}[t]
\centering
\includegraphics[width=\linewidth]{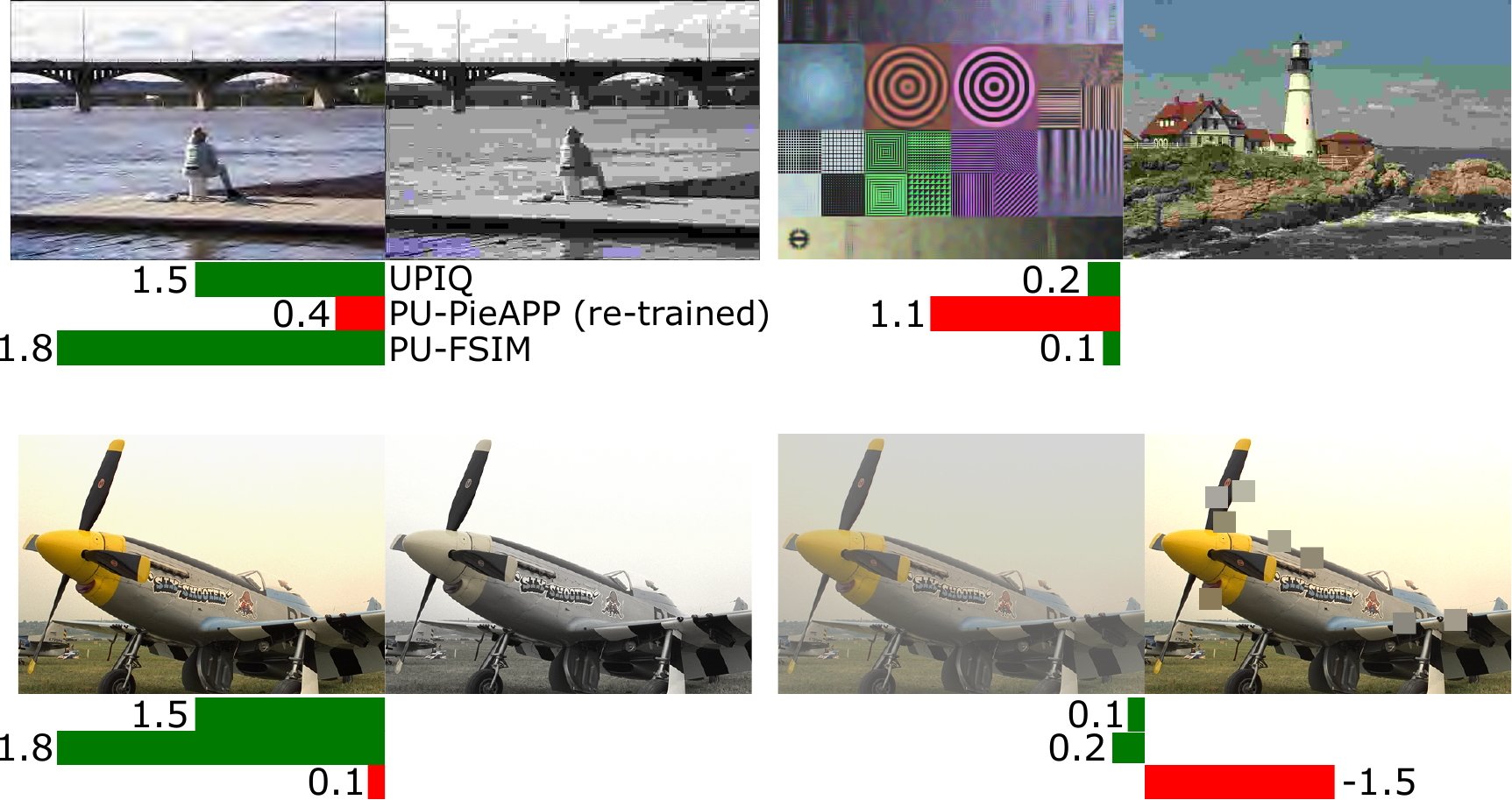}
    \caption{{Selection of image pairs from the gMAD competition where PU-PieAPP (re-trained) and PU-FSIM disagree the most. In the top row we show successful attack and successful defence of the PU-FSIM and in the bottom row the same for PU-PieAPP (re-trained). Below each pair we report the difference in the JOD scores between the left and the right image found in the UPIQ dataset and assigned by each of the tested metrics.}} \label{fig:mad_metrics}
\end{figure}

{We use the gMAD \cite{Ma2020} procedure to find the pairs of images that differ the most according to one metric, but are similar according to another metric. Figure \ref{fig:mad_metrics} shows a set of failure cases for the two best performing metrics (PU-PieAPP (re-trained) and PU-FSIM) when paired against each other in the gMAD competition. PU-PieAPP tends to correctly capture the ranking of the images, however underestimates the quality of images with JPEG artifacts. PU-FSIM fails to account for color change of the image. PU-FSIM is also too sensitive to contrast change and not sensitive enough to the structural distortions.}

\subsection{Brightness-Adaptive Coding}\label{sec:br_adaptive_coding}

As HDR metrics account for absolute luminance levels, they are not only useful for testing the quality of HDR images, but also open opportunities for new applications such as brightness-adaptive coding. Such coding adapts the required bandwidth to the brightness of a display; saving bits on a dimmed displays and using higher bandwidth when higher quality is needed for a bright display. 
We investigate how PU-PieAPP and PU-FSIM (the two best-performing metrics) predict the quality of images shown on displays of different brightness. Figure~\ref{fig:pred_across_lum} shows the quality predictions for JPEG distortion as a function of peak display luminance. As expected, both metrics predict that JPEG distortions are less noticeable when the display is darker. 
However, the predictions diverge at luminance levels above 100\cdms: PU-FSIM predicts a decrease in image quality, whereas PU-PieAPP predicts an improvement. Interestingly, PU-PieAPP's U-shaped curve is consistent with the recent measurements \cite{Wuerger2020} of human contrast detection thresholds. We show an example of these measurements in  Figure~\ref{fig:pred_across_lum}.

Next, we use PU-PieApp to control the compression rate of a standard JPEG codec (the "quality" parameter) in order to achieve a distortion at a desirable JOD level. Figure~\ref{fig:application} shows the distribution of the required bit-rate to compress 200 pristine test images at the desired JOD level. The selected levels signify that about 4\% (-0.1~JOD), 16\% (-0.3~JOD) or 25\% (-0.5~JOD) will correctly indicate a compressed image from a test and reference pair (discounting 50\% guess rate). The vertical bars in the plot denote the peak luminance levels of three typical displays: an HDR~TV, computer monitor, and a dimmed mobile phone.  The plot shows that the bit-rate could be substantially reduced when images are shown on a dimmed mobile phone, but it should be increased for HDR TV. Furthermore, the difference between brightness levels is larger for images encoded with high quality. Such information could be useful, for example, for internet caches that attempt to reduce the amount of data sent to mobile web browsers. It is only meaningful to use photometric metrics, trained on both SDR and HDR images, for such applications as they can capture the effect of absolute luminance on image quality. A more detailed validation of such brightness-adaptive image coding can be found in \cite{ye2020brightness_metrics}.

\section{Conclusions}
A large scale photometric image quality dataset would enable the development of deep learning based image quality metric. However, existing HDR image quality datasets are small in size and expensive to collect. We remedy this limitation and increase their size by merging together a mixture of both HDR and SDR datasets. Our merging procedure requires collecting additional data (cross-dataset comparisons), however, the experimental effort is much smaller compared to collecting the dataset from scratch. The accuracy of the resulting dataset is much higher than that of alternative procedures \cite{INLSA,Zerman2017_allignment}. The proposed dataset merging procedure can be applied to other quality domains, such as the quality of high-frame-rate, omni-directional or foveated video. 

Another major contribution of this work is Unified Photometric Image Quality dataset (UPIQ), which is the first large-scale dataset that can be used for training and testing HDR image quality metrics. Images in our dataset are represented in absolute photometric and colorimetric units and their quality scores are represented in the interpretable JOD units \cite{2019TIP}. We use the dataset to (a) train a deep-leaning based quality metric for HDR images (PU-PieApp); (b) benchmark the state-of-the-art HDR image quality metrics; and (c) demonstrate how trained quality metrics can be used for brightness-adaptive image coding. All those applications demonstrate the benefits of UPIQ and the need for photometric image quality metrics.

\section*{Acknowledgment}
This project has received funding from EPSRC research grants EP/P007902/1 and EP/R013616/1, from the European Research Council (ERC) under the European Union’s Horizon 2020 research and innovation programme (grant agreement N$^\circ$ 725253 (EyeCode), and from the Marie Sk{\l}odowska-Curie grant agreement N$^\circ$ 765911 (RealVision).

\ifCLASSOPTIONcaptionsoff
  \newpage
\fi



\bibliographystyle{IEEEtran}
\bibliography{IEEEabrv,./ms.bib}
\end{document}


%
\title{Supplementary for: ``Consolidated Dataset and Metrics for High-Dynamic-Range Image Quality"}
%
%

\author{Aliaksei~Mikhailiuk, Mar\'ia P\'erez-Ortiz, Dingcheng~Yue, Wilson Suen, and Rafa{\l} K. Mantiuk
\thanks{A. Mikhailiuk, D. Yue, W. Suen and R. Mantiuk are with the Department of Computer Science and Technology at the University of Cambridge (UK) (email: \{am2442, dy276, wss28, rkm38\}@cam.ac.uk).} 
\thanks{M. P\'erez-Ortiz is with the Department of Computer Science at the University College London (UK) (email: maria.perez@ucl.ac.uk)}
}

%
%

\markboth{Journal of \LaTeX\ Class Files,~Vol.~14, No.~8, August~2015}%
{Shell \MakeLowercase{\textit{et al.}}: Bare Demo of IEEEtran.cls for IEEE Journals}
%



\maketitle

\IEEEpeerreviewmaketitle

\setcounter{figure}{11}   

\section{Introduction}
\IEEEPARstart{T}{his} supplementary file contains additional information that we could not include in the main paper due to space considerations. Below we describe: (i) the experimental procedure for the cross-dataset and within dataset pairwise comparisons; (ii) justification for the linear complexity of the model for psychometric scaling; (iii) detailed architecture of the PU-PieApp; (iv) selection procedure for datasets included in  UPIQ (v) detailed architecture of the multitask network; (vi) Maximum Differentiation (MAD) competition for the tested metrics; (vii) example images from the UPIQ dataset.

\section{Experimental Protocol for Dataset Merging}\label{sec:ap_exp}

Here we include additional details about the design of the experiments for collecting required cross-dataset and within dataset quality measurements. In order to produce a meaningful unified quality scale using pairwise comparisons for a specific single IQA dataset one needs a) comparisons of distorted to pristine quality reference image, b) within-content comparisons to scale different levels of distortion for the same distortion type and c) cross-content comparisons \cite{Zerman2018}, to connect all content and put them on the same quality scale. For rating this would be equivalent to having observers rate images across all distortions and distortion levels during the same session, instead of having separate experiments. 
In the case of selected datasets, all of these considerations were taken into account when original data was collected, i.e. each dataset has a self-contained unified quality scale. To align these datasets we need to connect disjoint scales through pairwise comparisons and also find the relationship between rating and pairwise comparison judgments within each of the datasets. This means that for every disjoint rating dataset we need to collect within dataset comparisons and link all datasets with across dataset comparisons.

\subsection{Displays and stimuli} 

For the presentation on the HDR display we transform all images from either gamma-corrected (SDR) or relative linear (HDR) pixel value to absolute linear colorimetric units in the Rec. 709 colour space \cite{BT709}. The peak luminance of the images was matched to the peak luminance of the displays used to collect original datasets. The images were also displayed with the same angular resolution (in pixels per visual degree) as in the original experiments. When the image size exceeded the size of our display, we provided a simple panning interface in which observers could use a trackball to inspect different portion of the image.

\subsection{Experimental Procedure and Participants} 
We extended the data collected in original datasets and follow-up studies for TID2013 \cite{Mikhailiuk2018,2019TIP} and LIVE \cite{ye2014active} datasets with two additional pairwise comparison experiments. In all cases, comparisons to be performed were selected so that compared images were of similar quality, excluding obvious comparisons so as to maximise informativeness of the collected data. Note that this is a common approach in pairwise comparison experiments and the basis for active sampling approaches \cite{ye2014active}. 

In the first experiment we collected only comparisons within the dataset, \ie~ comparing images of the same dataset. This is necessary for finding the relationship between rating measurements and pairwise comparisons. It is only necessary for rating-based datasets, which means we excluded TID2013 from this experiment since we used previously collected pairwise comparisons and rating measurements \cite{Mikhailiuk2018,2019TIP}. We ensured that all three types of previously mentioned comparisons were covered: to reference, within-content and cross-content. After the first experiment, all the data could be scaled, since we had comparisons to a common reference for all datasets. 

For the second experiment we compared conditions exclusively from different datasets, connecting each dataset to the rest. Images were chosen to uniformly cover the quality scale. We performed several iterations of the pair selection. After conducting a pairwise comparison experiment on a small batch of comparisons, we re-scaled the dataset with newly collected comparisons and selected the next batch from the new scale.

Observers were asked to compare two distorted images and choose the one with better quality with respect to their reference. The reference image could be viewed by pressing and holding a space bar. The observer was asked to see the reference images at least once for each comparison. The order of comparisons in every experiment was randomized. We ensure that ITU recommendations \cite{BT500} were met. And that the time for performing one experiment did not exceed 30 minutes, so as to prevent observer tiredness from influencing the experiment outcomes. Each selected pair of images was compared by 6 participants, with each participant completing approximately 300 trials. Overall 6000 new comparisons were collected from 20 participants. 

\section{Relationship between pairwise comparisons and mean opinion scores} \label{sec:ap_relation}

Watson~\cite{Watson2001} studied the correlation between rating scales and results of pairwise comparisons in the context of psychometric scaling of pairwise preference probabilities. He found that the degree of agreement between two scales, for the case of video compression, is relatively high. The work reports a quadratic relationship between MOS and scaled PWC, with a very small quadratic coefficient. On the contrary \cite{Zerman2018} shows that there is a strong linear relationship between MOS and PWC scaling results. Here we test both assumptions to validate, if the linear relationship is indeed sufficient.

To compare goodness of fit we report adjusted $R^2$ statistic -- $R^2_{adj}$, which, unlike simple $R^2$ accounts for the number of model parameters in explaining the variance in the data \cite{murphy_2012}:

\begin{equation}
    R^2_{adj} = 1 - (1-R^2)\frac{(n-1)}{n-p-1},
\end{equation}

where $R^2$ is defined as:

\begin{equation}
    R^2 = \frac{\sum_i^n (y_i-\hat{y}_i)^2}{\sum_i^n (y_i-\Bar{y}_i)^2},
\end{equation}
where $n$ is the number of data points in the dataset, $p$ is the number of parameters, excluding the constant term, $\mathbf{y}$ and $\hat{\mathbf{y}}$ true and predicted response variables and $\Bar{y}$ is the mean of $\mathbf{y}$.

\begin{figure}[h!]
    \centering
      \subfloat[TID2013]{%
       \includegraphics[width=0.5\linewidth]{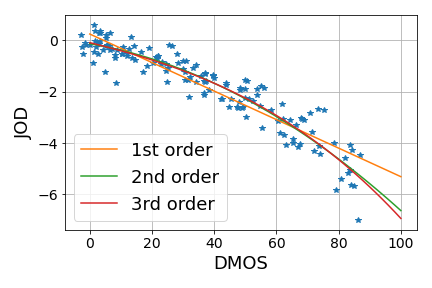}
       }
    \subfloat[LIVE]{%
       \includegraphics[width=0.5\linewidth]{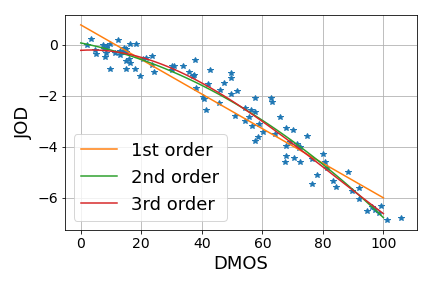}
       }
  \caption{Polynomial fits into the JOD and MOS scores of three subjective image and video quality datasets.}
  \label{fig:models_fit_complexity} 
\end{figure}

We fit $1^{st}$ $2^{nd}$ and $3^{rd}$ order polynomials into the JOD, obtained from pairwise comparisons, and MOS of TID2013 \cite{Ponomarenko2015} and LIVE \cite{Sheikh2006b} image quality datasets. Figure \ref{fig:models_fit_complexity} shows the scatter plot of the scores and fitted polynomials. An important observation can be drawn -- the model describing the relationship between JOD and MOS for image quality must be monotonic, as an increase/decrease in the quality of an image should result in the increase/decrease of the scores in both scales. Violation of this requirement is visible in the example of $3^{rd}$ order fit into the scores from LIVE dataset.

The results of computing $R^2_{adj}$ are given in Table  \ref{tb:rsq_model_complexity}. There is only slight increase in $R^2_{adj}$ for TID2013 and LIVE datasets for $2^{nd}$ and $3^{rd}$ order polynomials. Non-linear relationship is thus hard to justify given the need for additional constraints on the function to be monotonic.

\begin{table}[ht!]
\centering
\small
\caption{$R^2_{adj}$ statistic for polynomial fits describing the relationship between MOS and JOD.}
\begin{tabular}{| c | c | c | c |} 
\hline
	  Dataset & $1^{st}$ order & $2^{nd}$ order & $3^{rd}$ order\\
	  \hline
	  \hline
	  TID2013  & 0.77 & 0.79  & 0.79 \\
	  \hline
	  LIVE  & 0.87 & 0.89  & .89\\
\hline

\end{tabular}
  \label{tb:rsq_model_complexity} 
\end{table}

\section{UPIQ dataset selection}
\begin{figure}[t]
\centering
\includegraphics[width=.7\linewidth]{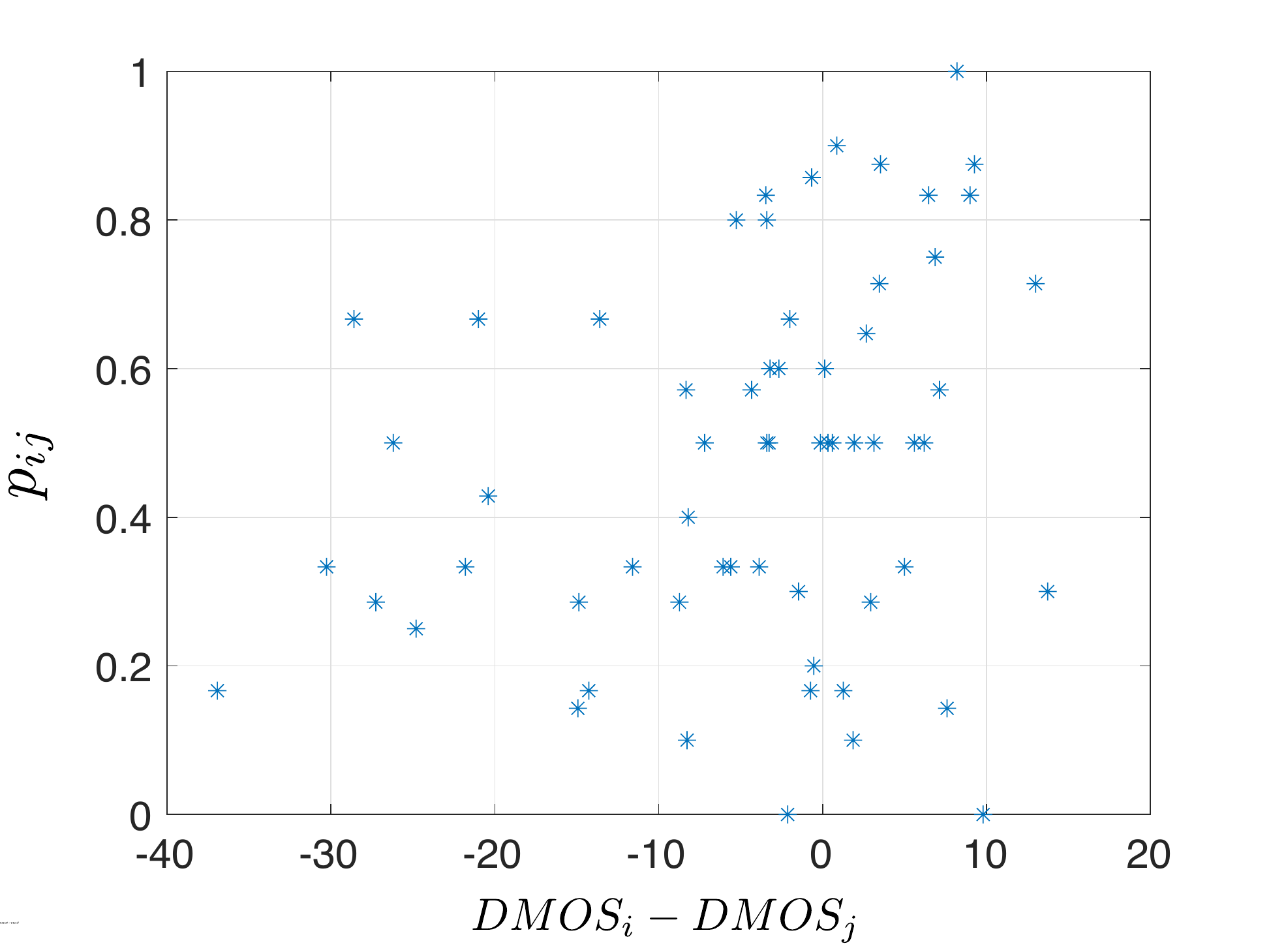}
    \caption{{Scatter plot of the empirical probability $p_{ij}$ obtained from our experiment and difference in the DMOS scores obtained by \cite{Zerman2017_allignment}. Two scales have little correlation with the SROCC of 0.27.}} \label{fig:scatter_plot_zerman_vs_prob}
\end{figure}

{To ensure the accuracy of the data in the UPIQ dataset, candidate datasets were screened with a pilot experiment. We ran a series of within-dataset pairwise comparisons to verify if the ranking of the scores elicited from our subjective study is consistent with the ranking provided in the dataset. Where the scores in the original dataset have shown to have little or no correlation with the data in our experiment, the dataset was not included in UPIQ. Figure \ref{fig:scatter_plot_zerman_vs_prob} shows the scatter plot of the empirical probabilities found in our subjective experiment for a set of image pairs, compared between six and ten times, versus their difference in the MOS scale obtained by \cite{Zerman2017_allignment}.}

\begin{figure}[t]
\centering
\includegraphics[width=\linewidth]{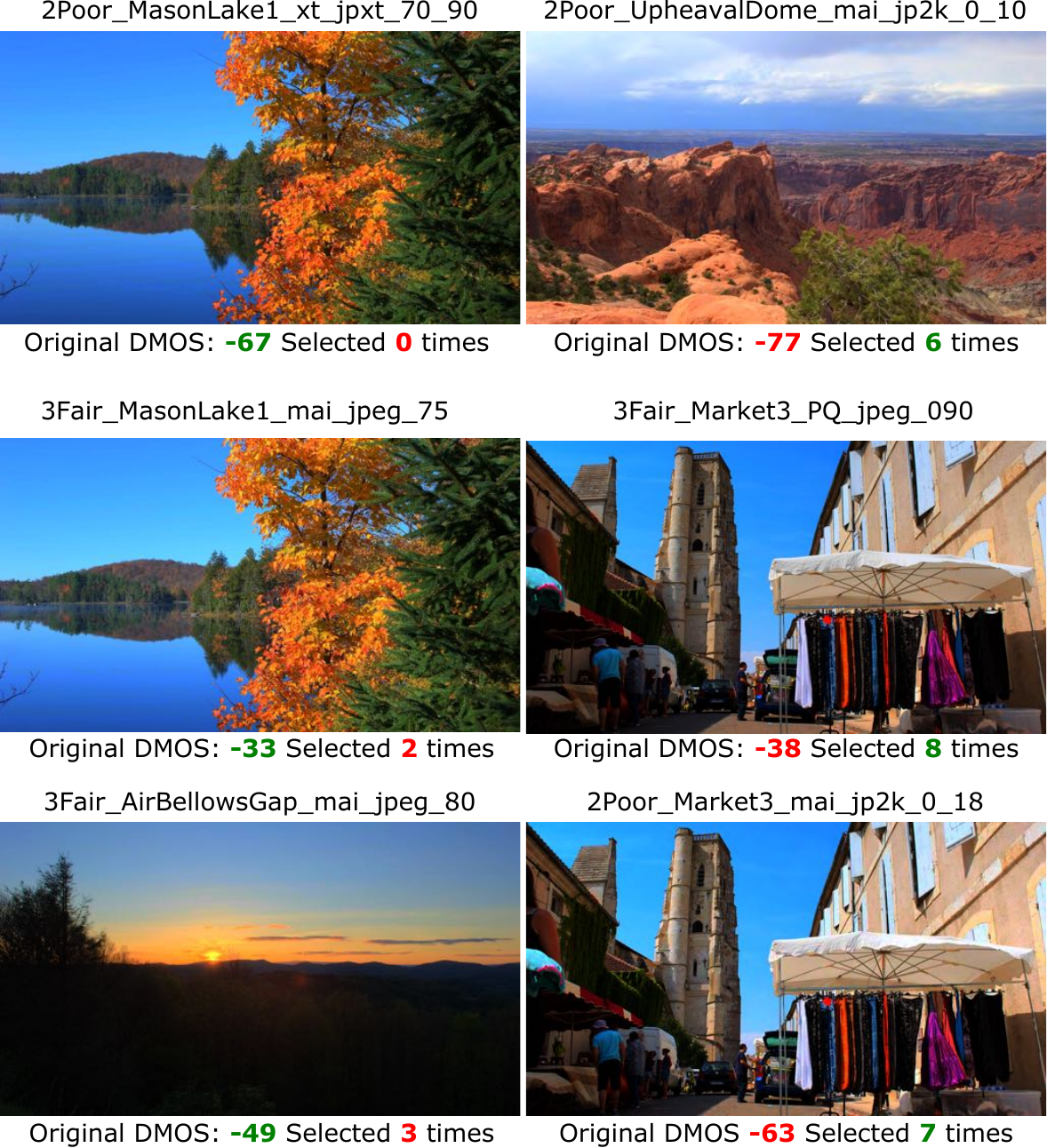}
    \caption{{Representation of image pairs where collected pairwise comparisons and original DMOS scores from \cite{Zerman2017_allignment} disagree. In each pair the image on the left is the one which has higher DMOS and lower quality from our experiment and the other way around on the right.}} \label{fig:zerman_mad_examples}
\end{figure}

{We verify if the data collected by \cite{Zerman2017_allignment} or us is more consistent we run an additional Maximum Differentiation (MAD) experiment \cite{zhou2008}. The pairs of images with the most inconsistent scores are given in Figure \ref{fig:zerman_mad_examples}. One of the advantages of the experimental procedure employed for the data collection in UPIQ is the ability to flip between reference and test image during the experiment. Observers were thus, particularly sensitive to the JPEG blocking artefacts in the large smooth areas of the skies of the lake and sunset images.}

\section{PieAPP Detailed Architecture}\label{sec:ap_pu_pieapp}

\begin{figure}[t]
\centering
    \begin{subfigure}[b]{0.5\textwidth}            
            \includegraphics[width=\linewidth]{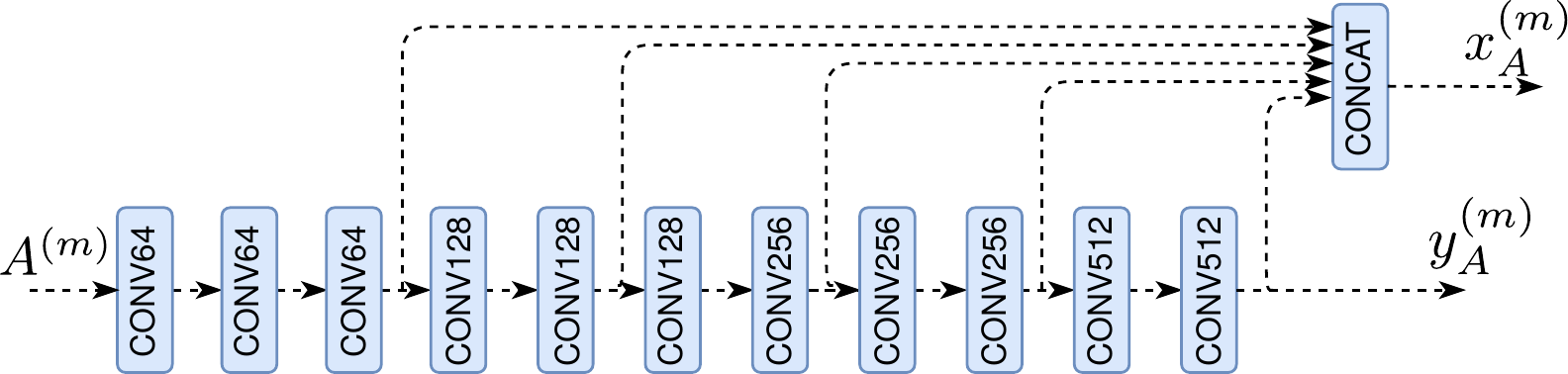}
            \caption{Feature extraction}
    \end{subfigure}%
    
    \begin{subfigure}[b]{0.5\textwidth}
            \centering
            \includegraphics[width=\linewidth]{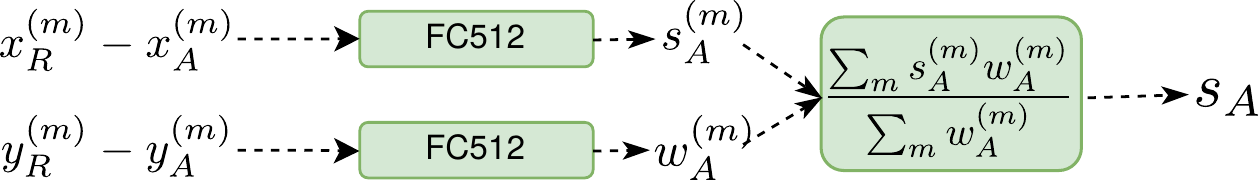}
            \caption{Score computation}
    \end{subfigure}
    \caption{(a) The feature extraction network takes image patches as an input and has two outputs: one from a patch passing through the whole network and another formed from skip connection. The network has 11 convolutional layers with $2\times2$ max-pooling after every even layer. (b) The score computation network computes patch-wise weights and scores, the weighted average produces the final score}\label{fig:pieapp_branches}
\end{figure}

For every input patch $m$ of reference $R$ and distorted $A$ images the feature extraction (FE) network has two outputs: $y^{(m)}$ from the input passing through the whole network and $x^{(m)}$ formed by concatenation of the flattened outputs of layers at different depths of the network. The score computation (SC) network takes two inputs: the difference between $x^{(m)}_R - x^{(m)}_A$, which is passed through a fully connected layer, predicting patch-wise error $s^m$ and the difference $y^{(m)}_R - y^{(m)}_A$, which is passed through another fully connected layer, producing the patch-wise weight $w^{(m)}$. The two outputs $s^{(m)}$ and $w^{(m)}$, are then used to produce the weighted average of all per patch scores -- a quality score of the entire image $s_A$. Note, that passing two reference images through the network will result in the $x^{(m)}_R - x^{(m)}_A = 0$, thus the output of the quality estimation function $f(A,B)$, will be constant, defined by the bias of the score computation network. The detailed architecture of the PieAPP network is shown in Figure \ref{fig:pieapp_branches}. 

\paragraph{Alternative Architectures} We experiment with a number of CNN architectures to find the one that generalizes the best. Since the CNN-based metric can be trained end-to-end, it could potentially learn the PU-transform. We replaced the PU-transform with a logarithmic function followed by scaling to the 0-1 range and then trained the network. The prediction error was much higher for the logarithmic function (\ac{RMSE} 0.68) compared to the PU-transform (\ac{RMSE} 0.47). This confirms that the PU is beneficial for quality predictions in SDR/HDR images even for CNN-based metrics.

\section{Multitask Network}\label{sec:ap_mt}

\begin{table}[!t]
    \small
    \renewcommand{\arraystretch}{1}
    \centering
    \begin{tabular}{|c|c|c|c|c|c|}
        \hline 
        \makecell{TID2013\\LIVE} &
         \makecell{Narw.\\Korsh.} & 
         \makecell{TID2013\\Narw.} & 
         \makecell{LIVE\\Narw.} & 
         \makecell{TID2013\\Korsh.} & 
         \makecell{LIVE\\Korsh.} \\
         \hline
         0.46 & 0.27 &  0.33 & 0.46 & 0.26 & 0.10 \\
        \hline

    \end{tabular}
    \caption{SROCC between the difference in quality scores $s_A-s_B$, where $A$ and $B$ are images from different datasets and empirical probability $p_{ij}$ for the multitask network.}
    \label{tb:srocc-mt}
\end{table}

\begin{figure}[t]
\centering
\includegraphics[width=\linewidth]{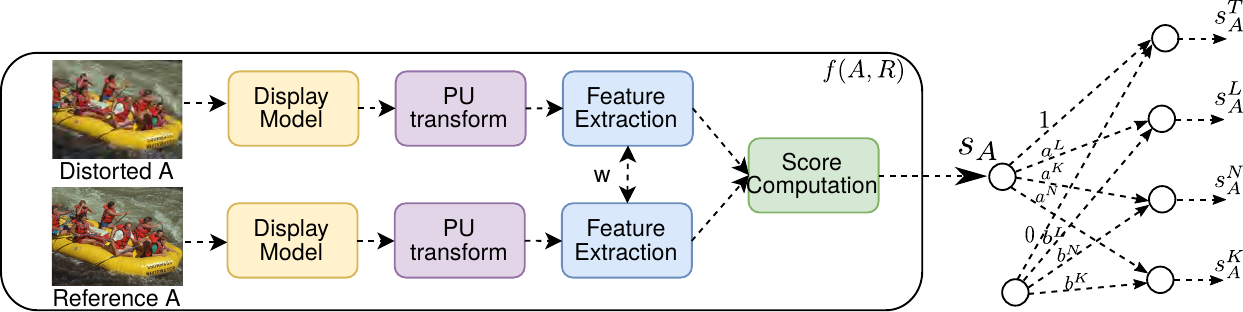}
    \caption{Multitask network. The network is trained to predict original scores from individual datasets. Similar to the scaling procedure the network learns the implicit quality $s_A$ and parameters $a$ and $b$ for each dataset. To constraint the scores we set parameters of TID dataset $a^T=1$, $b^T=0$.} \label{fig:mtask_net}
\end{figure}

Collecting data is time consuming and expensive, hence a method capable of learning the implicit unified quality without the need for additional data is desirable. To verify if our network is capable of learning this implicit quality and cross-dataset relationship, we train the network using a multitask learning approach, where it is assumed that all datasets share the same feature representation for quality but since scales are relative the quality scores might be scaled differently. The architecture of the network is given in Figure \ref{fig:mtask_net}. The $f(A,B)$ part of the network is the same as PU-PieApp and produces a score $s_A$, which is assumed to be a unified quality for disjoint datasets. Similar to our scaling procedure from Section 3 of the main paper the scores from individual datasets are linked with the unified $s_A$ via a linear relationship. For example the quality score $s_A^L$ for LIVE dataset would be predicted with $a^L*s_A+ b^L$, where $a^L$ and $b^L$ are learnt parameters. These parameters from individual datasets are treated and learnt as individual tasks. Since quality scores are relative, we constraint them by setting parameters of TID2013 dataset $a^T=1$ and $b^T=0$. To allow for faster convergence we standardized scores from the separate datasets. The training procedure for the multitask network was the same as for the DPIQM. 

Similar to Section 4.3 in the main paper we compute the correlation between the difference in quality scores $s_A-s_B$, where $A$ and $B$ are images from different datasets and empirical probability $p_{ij}$. The detailed results are given in Table \ref{tb:srocc-mt}. Neither of the cross-dataset relationships is well captured by the multitask network. 

\section{Maximum differentiation competition}

\begin{figure}[t]
\begin{center}
   \includegraphics[width=\linewidth]{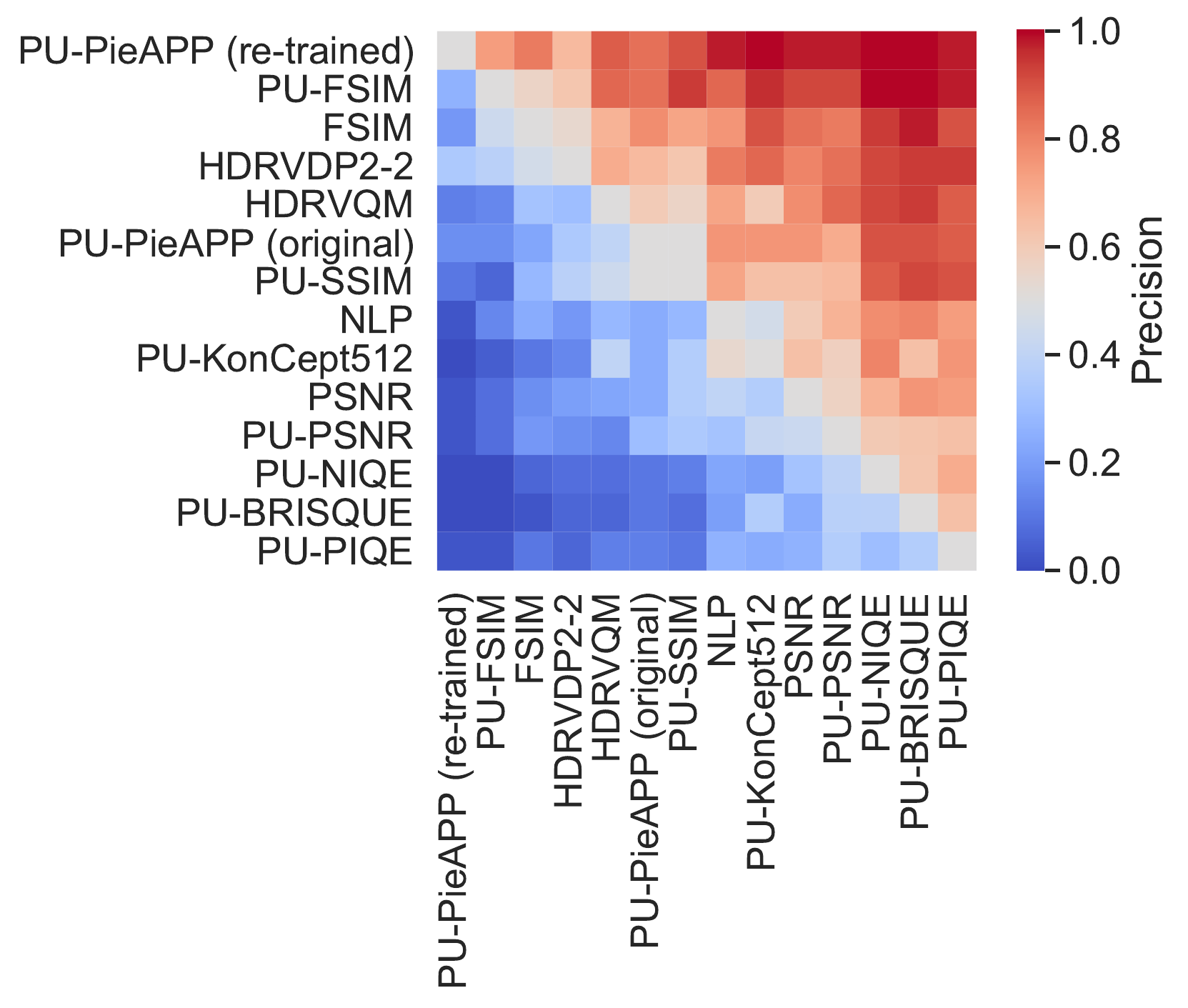}
\end{center}
\caption{{MAD competition for tested metrics. High values in the row indicate high success of the attack by the metric in the corresponding row on the metric in the corresponding column.}}
\label{fig:mad_competition}
\end{figure}

{We have also performed MAD analysis on the test split of the UPIQ dataset. For a pair of metrics we select pairs of conditions that have different qualities according to the tested quality metric and similar according to the benchmark quality metric. Thus for two quality metrics $M^t$ and $M^b$ with scores  in JOD units we select conditions $o_i$ and $o_j$ following $\argmax_{ij}(|M^t_i-M^t_j|-|M^b_i-M^b_j|)\ subject\ to\ |M^b_i-M^b_j|<1\ JOD$. Instead of aggressiveness and resistance used in \cite{Ma2020}, we quantify the performance of a metric by measuring its ability to classify a pair of images as of the same or of different quality. If the absolute difference in JOD units between two images in the UPIQ dataset is $<1$, we assume that the conditions are similar in quality, otherwise they are different in quality. We then report precision - the number of pairs correctly ranked and identified as different by the tested quality metric, divided by the total number of selected pairs (100 in our case). The results are given in Figure \ref{fig:mad_competition}. Each entry of the matrix is the precision of the test metric from the corresponding row when paired against the benchmark quality metric from the corresponding column.} 

{PU-Pie-APP (re-trained) exhibits the best performance both in identifying different in quality (first horizontal row) and similar in quality (first vertical column) conditions when paired with any of the metrics. However, when paired with HDRVDP2-2, PU-PieAPP performs almost on par. Second best performance is attained by PU-FSIM, which has very similar performance to FSIM without PU-transform. Nevertheless, PU-FSIM exhibits stronger performance when paired with metrics accounting for the dynamic range.}

\section{Examples of the dataset}\label{sec:ap_dataset}
\begin{figure*}[t]
\centering
\includegraphics[width=\linewidth]{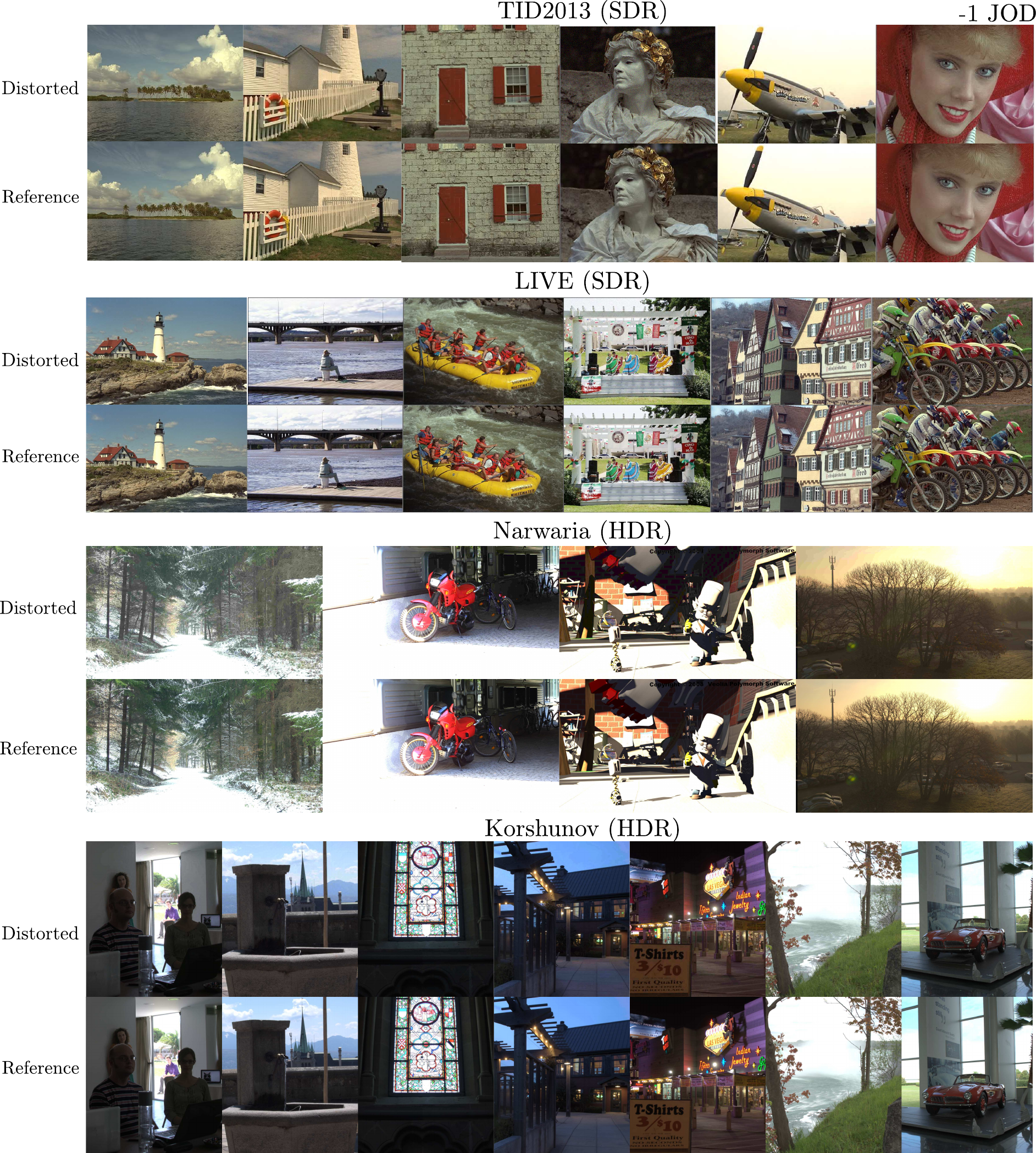}
    \caption{A selection of images from the four combined datasets at approximately $-1$ JOD level. Each dataset has two rows: distorted and reference images. We converted HDR images to SDR with gamma correction and gamma $2.2$. Images from different datasets at the same JOD level have similar distortion severity. Without a unified dataset it would be impossible to compare image scores across datasets.} \label{fig:minus_one}
\end{figure*}

\begin{figure*}[t]
\centering
\includegraphics[width=\linewidth]{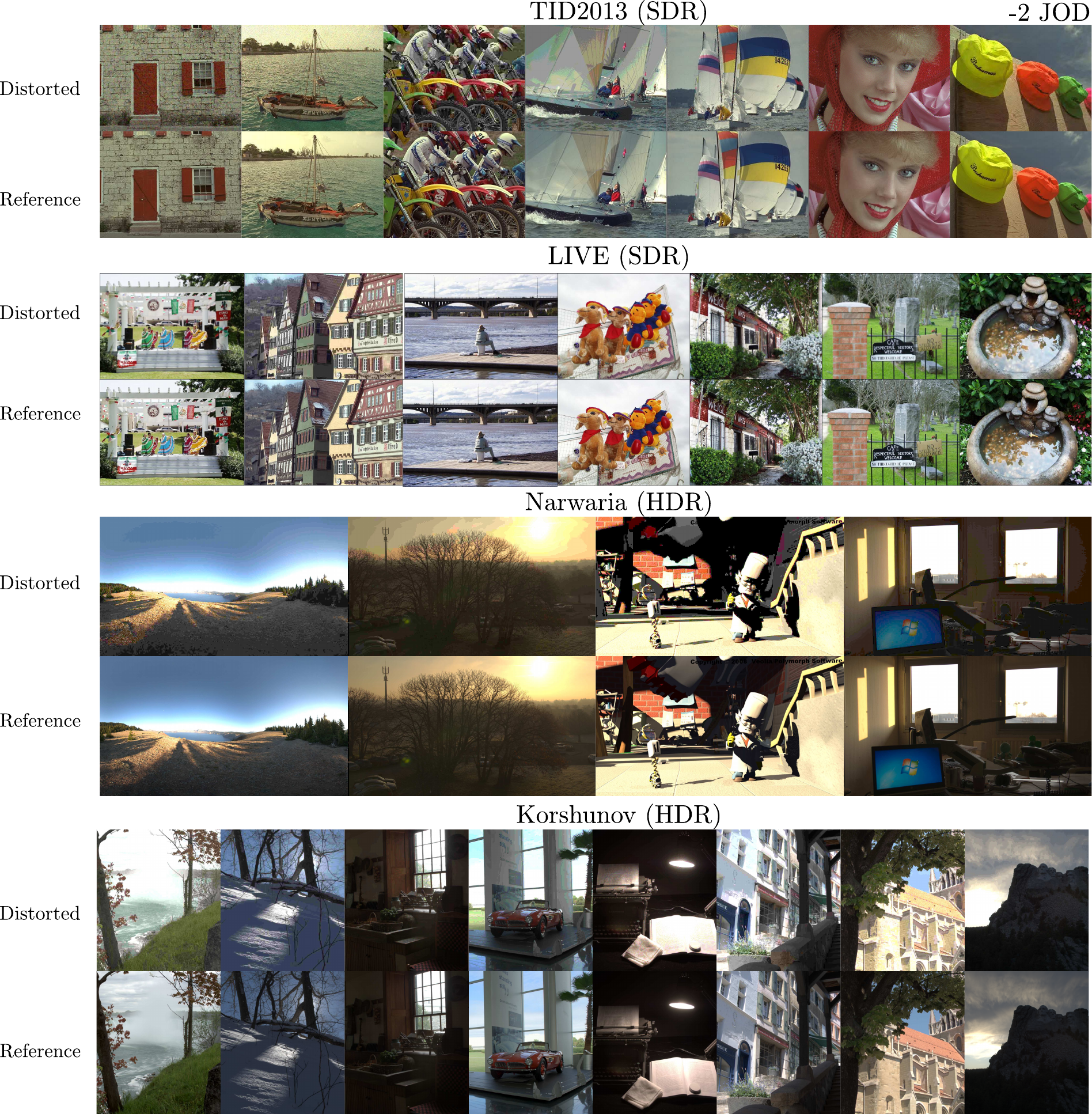}
    \caption{A selection of images from the four combined datasets at approximately $-2$ JOD level. Each dataset has two rows: distorted and reference images. We converted HDR images to SDR with gamma correction and gamma $2.2$. Images from different datasets at the same JOD level have similar distortion severity. Without a unified dataset it would be impossible to compare image scores across datasets.} \label{fig:minus_two}
\end{figure*}

Figures \ref{fig:minus_one} and \ref{fig:minus_two} show sample images from the unified dataset at $JOD = -1$ and $-2$ and are intended to be a visual subjective validation of the final scale. These levels were selected to show images from all four datasets, as images from the HDR datasets (Korshunov and Narwaria) have quality scores above -2 JOD only. Each figure contains four separated sections, each associated to a different dataset. Each section has two rows: distorted and reference images. For display purposes HDR images were converted to SDR with gamma correction:
\begin{equation}
    I_{\mathrm{HDR}} = 255 \left(\frac{ I_{\mathrm{SDR}}}{255}\right)^{\frac{1}{2.2}}.
\end{equation}
As the perceived image quality depends on the display luminance, the SDR images in the figures might be masking or amplifying some image distortions. Thus figures are intended to be an approximate demonstration of the final image quality scale. Nevertheless, images from different datasets at the same JOD level have similar distortion severity. Without our unified photometric image quality dataset (UPIQ) it would be impossible to compare image scores across datasets. Most of the HDR images are distorted only locally, with the overall image quality not deteriorating significantly, as opposed to images from SDR datasets that had uniform distortions applied to them. Narwaria mostly has panorama images, where local distortions are less noticeable due to the size of the image.

\ifCLASSOPTIONcaptionsoff
  \newpage
\fi



\bibliographystyle{IEEEtran}
\bibliography{IEEEabrv,./supplement.bib}